\documentclass[authoryear,preprint,3p,10pt]{elsarticle}

\usepackage{graphicx}%
\usepackage{multirow}%
\usepackage{amsmath,amssymb,amsfonts}%
\usepackage{amsthm}%
\usepackage{mathrsfs}%
\usepackage{mathtools}
\usepackage[title]{appendix}%
\usepackage{xcolor}%
\usepackage{textcomp}%
\usepackage{manyfoot}%
\usepackage{booktabs}%
\usepackage{algorithm}%
\usepackage{algorithmicx}%
\usepackage{algpseudocode}%
\usepackage{listings}%
\usepackage{amsmath,bm}
\usepackage{tabularx}
\usepackage{ulem}
\usepackage[utf8]{inputenc}

\usepackage{lineno}
\usepackage{multirow}
\usepackage{colortbl} 
\usepackage[colorlinks=true,linkcolor=blue]{hyperref}%
\usepackage{soul}
\usepackage{comment}

\journal{Journal of Fluids and Structures}

\begin{document}

\begin{frontmatter}



\title{Self-adaptive elastic flaps with bending and torsion for 3D blunt body drag reduction}

\author[inst1,inst2]{J. M. Camacho-Sánchez}
\affiliation[inst1]{organization={Department of Mechanical and Mining Engineering, University of Jaén}, country={Spain}}
\affiliation[inst2]{organization={Andalusian Institute for Earth System Research, Universities of Granada, Jaén and Córdoba}, country={Spain}}

\author[inst2,inst3]{M. Lorite-Díez \corref{cor1}}
\affiliation[inst3]{organization={Department of Structural Mechanics and Hydraulic Engineering, University of Granada}, country={Spain}}
\cortext[cor1]{Corresponding author: M. Lorite-Díez, email: mldiez@ugr.es} % Update with actual email

\author[inst4]{Y. Fan}
\affiliation[inst4]{organization={Department of Aerospace Engineering, University of Liverpool}, country={UK}}

\author[inst2,inst3]{J. I. Jiménez-González}
\author[inst4]{O. Cadot}

\begin{abstract}
This study investigates the potential for drag-reduction of low-mechanical-order, self-adaptive control systems, consisting of hinged flaps attached along the edges of the rectangular base of a canonical blunt body. Comparative experiments are conducted in a wind tunnel under crosswind conditions at a Reynolds number of $Re = 2.13 \times 10^5$. The flaps, made of rigid rectangular panels, are mounted in three configurations: rigidly fixed%(RF)
, flexibly hinged with a single degree of freedom in bending%(1HF)
, and flexibly hinged with two degrees of freedom, allowing both bending and torsion%(2HF)
, the latter representing a novel drag-reduction device, easily tunable to ensure a quasi-steady, stable adaptive reconfiguration. 
Two geometric arrangements are tested: horizontal flaps attached along the top and bottom (TB) edges, and vertical flaps along the lateral (left and right, LR) edges. The experimental study includes force, pressure, flap deformation and wake velocity measurements at varying yaw angles to simulate crosswind conditions.  When the body is aligned with the flow, both arrangements reduce drag due to a rear cavity effect that elongates the recirculating flow. The TB arrangement is found to be much more effective at reducing drag in yawed conditions and its performance is improved using the flexible hinges. In these cases, static deformations correspond to boat-tailing that reduces the induced drag together with the turbulent kinetic energy in the wake. The use of the wind-average drag coefficient (taking into account events of crosswind) to evaluate an effective drag reduction clearly shows the TB arrangement with bending and torsion as the best appendage, with a 7.62\% drag reduction compared to the body with no appendages, proving the good performance of simple, two-degrees-of-freedom control systems to adapt to changing three-dimensional wakes.

\end{abstract}


% \begin{highlights}
% \end{highlights}


\end{frontmatter}
%\linenumbers
%\tcr{Main changes: subscript $_f$ for flap and never for fluid- dissertation in a present/passive form, never in the past form unless necessary. $f^*$ for dimensionless frequency as defined in the text instead of St.}
\section{Introduction}
\label{sec:Introduction}
%% Drag reduction in road transport, CO2, off design conditions
The sustainable development goals proposed by United Nations include reducing the environmental impact of the transport industry \citep{UN2016}. According to the European Environment Agency,  greenhouse gas emissions from the transportation sector must be reduced by 30$\%$ by 2030 \citep{EEA21}. Medium- and heavy-duty vehicles account for a significant proportion of the global energy consumption of the transport industry, and the electric versions of these vehicles are often not viable under real conditions because of the large energy-storage requirements. These vehicles typically operate at a constant speed on motorways or highways, where aerodynamic losses can account for up to 80$\%$ of the total energy consumption \citep{NRC2010}.  

The design of trucks and buses has traditionally been driven by functional considerations, including maximizing the load capacity, ensuring compatibility with loading docks, and enhancing driver comfort. However, aerodynamic performance has often been overlooked during the design phase despite its potential impact on fuel efficiency and overall vehicle performance. Therefore, there is plenty of room to optimize their aerodynamics. The drag of these vehicles is generated in different zones; in particular, the base of the vehicle is responsible for approximately 25$\%$ of the total drag owing to the low pressure generated by the rear massive flow separation induced by the blunt shape \citep{Wood03}. The flow separation creates a fully three-dimensional, turbulent, and recirculating wake, which increases the drag of the heavy vehicle and, therefore, its energy consumption. The aerodynamic drag  also changes under different flow conditions such as crosswinds \citep{Fan2022}, gusts \citep{Kozmar2012}, platooning \citep{Robertson2019} or ambient turbulence \citep{Passaggia2021}. 

An additional source of aerodynamic forcing in near wake of blunt bodies is the presence of the Reflectional Symmetry Breaking (RSB) mode, in which the wake randomly switches between two horizontally deflected mirror positions, both in simplified models \citep{Grandemange13b,Volpe2015} and in full-scale realistic configurations \citep{Bonnavion2019b}.  It has also been demonstrated that this bistable behaviour is highly sensitive to the Reynolds number \citep{Fan2023}, the alignment of the model with the flow (pitch and yaw angles) \citep{Cadot15,Fan2022}, the model aspect ratio \citep{Grandemange13a}, and several other factors. This bistable dynamics has been shown to affect the overall drag of the model \citep{Haffner20}, due to the increased interaction of the shear layers during the occurrence of each of these states.
Consequently, mitigation strategies have been investigated throughout the literature.

%Intensive research has been devoted to the development of control devices in the rear parts to improve aerodynamic performancse \citep{Choi14}. The research is usually focused on simplified models of heavy vehicles consisting in canonical three-dimensional blunt bodies, such as the square-back Ahmed body \citep{Ahmed84}, which allows to study in a fundamental way the different control strategies proposed and to gain a physical understanding of their effect , which is the basis for future developments more related to heavy vehicles and real operating conditions \citep{Choi14}. 

%% Passive control (flaps)
%% Bio inspited Self-adaptive solutions: unsteady or off desing conditions
In that context, passive control devices have been proven as efficient, simple alternatives to control the wake behind three-dimensional blunt bodies and reduce the drag. These strategies typically act by modifying the flow detachment or rear geometry of models. Typical examples of rear passive devices in the literature are: boat tailings \citep{Wood03, Fan_2024_Expif}, cavities \citep{Mair1978, Khalighi01, Evrard16, Lorite20b, Darabasz_2023}, and flaps \citep{Grandemange_2013-ExiF, Delacruz17b, Urquhart2018}. The aerodynamic performance of these rigid devices is closely related to the design conditions, and they might have a detrimental effect under crosswind and transient conditions \citep[see e.g.][]{Lorite20b}, which are usually encountered by heavy vehicles in real driving \citep{Hucho93, DHooge14}. 

Given these limitations, the design of self-adaptive devices, based on reconfiguration principles of biological systems \citep[see e.g.][]{Delangre08, Gosselin10, Mazellier12}, may represent appealing alternatives. Experimental studies on the use of flexibly-hinged rotary flaps \citep{Garcia23} and elastic flaps \citep{Garcia21} for slender blunt-based bodies, demonstrated that these adaptive solutions may outperform static, rigid flaps under yawed and transient flow conditions, on account of their ability to adapt to changing wake features. These results have also been replicated for three-dimensional wakes behind simplified models of blunt vehicles  \citep{CamachoSanchez23, MunozHervas2024}, where flaps are also shown to symmetrise the wake, leading to a weakening of both side and lift forces. Additionally, the moving parts can  passively interact with 3D flow structures, as the Reflectional Symmetry Breaking (RSB) mode. % that dominate the quasi-steady topology of 3D wakes \citep[see e.g.][]{Grandemange2012, Grandemange13a} and creates additional drag \citep{Bonnavion18, Haffner20}.
In spite of the positive effect of these flexible systems, the mechanical properties of any flexible device used as a control instrument must be carefully chosen to ensure a quasi-steady response of the flaps, since a significant vibrating amplitude can be detrimental for aerodynamic or structural purposes, as shown in \citet{MunozHervas2024}.

In view of that, low-mechanical-order adaptive systems with limited degrees of freedom offer a promising strategy because their dynamics are easier to tune. However, the complexity of three-dimensional wakes suggests that simple rotary flaps might not be optimal solutions in terms of drag reduction and wake control, as the adaption is purely two-dimensional along the axis of rotation. That said, the present study aims to analyze the combined effect of rear systems with two rotary, i.e. bending and torsional, degrees of freedom, and its capacity to adapt to three-dimensional flow structures behind a simplified model of heavy vehicles in variable crosswinds. The dynamic behaviour, deflection of flaps and wake changes of different adaptive systems are investigated to determine whether the aerodynamic improvement is directly attributable to aerodynamic shaping or broader unsteady effects.  

%\tcr{To revise: In particular, a key focus of this study is to identify the optimal spatial arrangement of these self-adaptive flaps to minimize drag in variable crosswinds.% by interacting with the turbulent structures that appear in the flow around simplified models of heavy vehicles under yaw
%The comparison between flap arrangements also considers critical mechanical parameters such as the stiffness of the system and the degrees of freedom for the flap deformation. For this purpose, a novel flap able to twist and bend to better adapt to crosswind conditions is developed and tested. The relationship between drag reduction and flap deformation is investigated to determine whether this performance gain is directly attributable to aerodynamic shaping or broader unsteady effects.}
%In addition, we analyze the unsteady effects induced by the flap motion and the dominant response (bending versus torsion) governing the flap response to varying flow conditions. A coupled fluid-structure interaction (FSI) study is performed to highlight the interplay between flap dynamics and wake behaviour, and to shed light on how the wake and flow frequencies are synchronised with torsional and bending motions. By answering these fundamental questions, this study advances our understanding of the dynamic role of rotary flaps in improving the aerodynamics of canonical three-dimensional blunt bodies and provides insights into their optimal design and use for heavy-duty vehicles in real-world conditions.
%% Questions to answer in the paper, organization

The paper is organised as follows. Experimental details comprising, model geometry and wind tunnel descriptions, flap geometry and mechanic properties, as well as the measurements equipment are provided in \S \ref{sec:Exp}. Results are presented and discussed in  \S\ref{sec:Results} within 3 parts; the baseline with no flaps  (\S\ref{sec:baseline}), the flaps arrangement either along the top and bottom edge or along the left and right edge of the rectangular base (\S\ref{sec:arrangements}) and the effect of increasing the degrees of freedom of the elastic flaps (\S\ref{sec:DoFs}). Finally, the main conclusions are presented in  \S\ref{sec:Conclusions}.

\section{Experimental details}
\label{sec:Exp}
%% Set-up: Body, Wind tunnel, experimental conditions
%% Tested Configurations
%% Mechanical characterization
%% Measurements: Force, Pressure, deformation (laser and camera), Stereo PIV
\begin{figure}[t]
\centering
\includegraphics[width=0.8\textwidth]{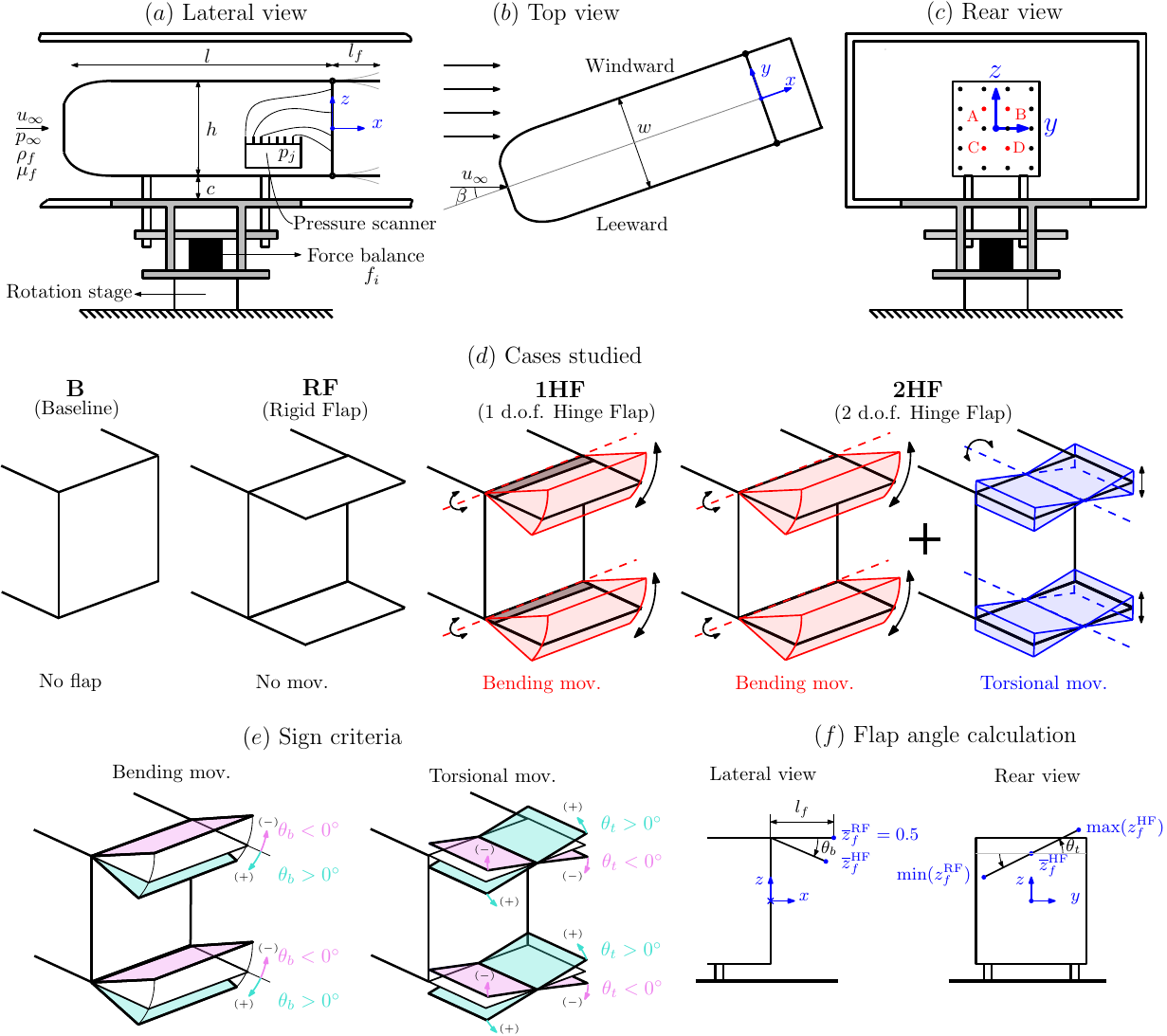}
\caption{\label{fig:SetUp}  $(a)$ Lateral, $(b)$ top,  and $(c)$ rear views of the squareback Ahmed body inside the wind tunnel. $(d)$ Sketch of the tested configurations for the Top-Bottom (TB) flap arrangement. $(e)$ Sign criteria used for the calculation of deflection angles. $(f)$ Sketch of the flap position for the determination of bending, $\theta_b$, and torsion, $\theta_t$, angles.}
\end{figure}

\begin{figure}[t]
\centering
\includegraphics[width=0.6\textwidth]{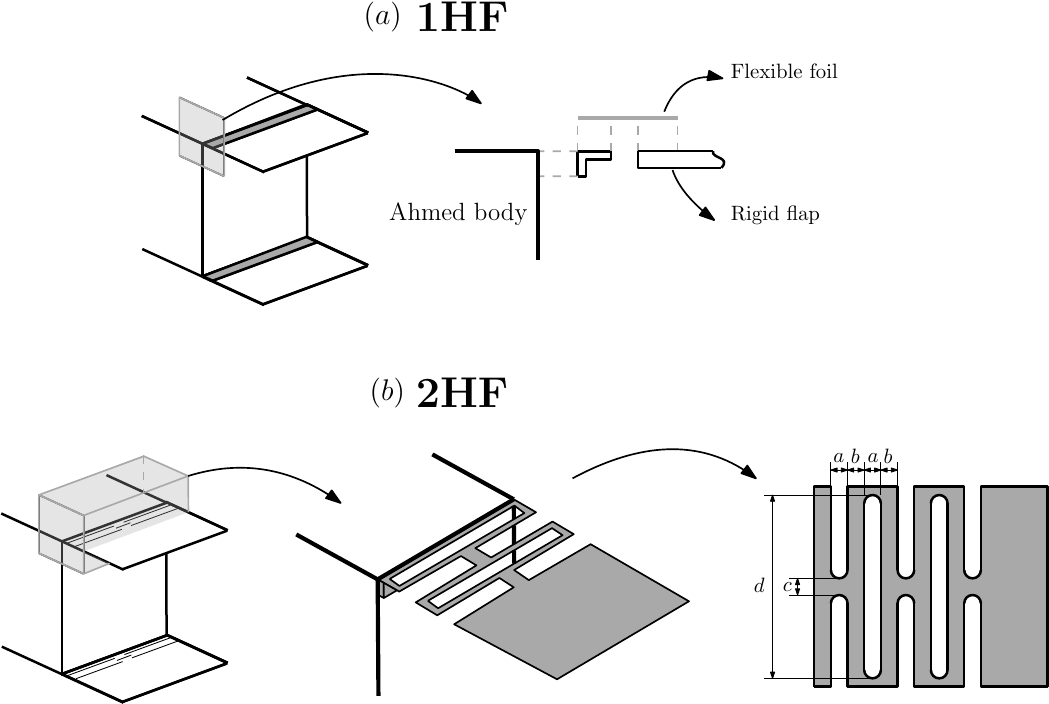}
\caption{\label{fig:FlapDetailed} Detailed sketch of the flexibly hinged systems for $(a)$ 1HF and $(b)$ 2HF configurations. In $(b)$ a Triple Lamina Emergent Torsional (LET) joint is used to confer an elastic deformation in bending as in $(a)$ and an additional elastic deformation in torsion.}
\end{figure}

\subsection{Model and wind tunnel}
\label{subsec:Setup}
%% Body
The model is based on the geometry originally introduced by \cite{Ahmed84}, which has become a benchmark for simplified road vehicle shapes. As shown in Fig.~\ref{fig:SetUp}, it has a height of $h = 0.2$~m, a width of $w = 0.18$~m ($w/h = 0.9$), and a total length of $l = 0.56$~m ($l/h = 2.8$). The forebody features a rounded leading edge with a radius of 0.07~m. The body is supported by four cylindrical supports with a diameter of 15~mm, positioned at a fixed ground clearance of $c/h=0.15$. The model is mounted on a motorised rotation stage (Standa 8MR190-90-59) to control the yaw angle $\beta$ (see Fig.~\ref{fig:SetUp}($b$) with a precision of $0.02^{\circ}$. The range of yaw angles that is investigated during the experiments is $\beta \in [-15:1:15]^\circ$.

%% Wind tunnel
The experiments are conducted in an open-circuit blow-down facility. %at the University of Liverpool. 
The airflow is driven by a fan located upstream of a contraction section with a 9:1 area ratio, followed by a rectangular test section. The test section measures 1.2~m in width, 0.6~m in height, and 2.4~m in length. The blockage ratio varies from $5\%$ to $9\%$ as the yaw angle $\beta$ increases from $0^{\circ}$ to $15^{\circ}$. The turbulence intensity in the streamwise direction, at the tested wind tunnel velocities, is approximately $0.2\%$. The usual free-stream velocity is $u_{\infty}= 16$ ~m/s, which sets the Reynolds number to $Re  = \rho u_{\infty}h/\mu = 2.13 \cdot 10^5$, where $\rho$ and $\mu$ are respectively the incoming fluid density and dynamic viscosity. The distance between the model and the test section inlet is 535~mm, such that the boundary layer thickness on the ground in front of the body is $\delta_{99}=7.5$~mm when the test section is empty. Additionally, lower wind tunnel velocities are tested to characterize the fluid structure interaction of the elastic drag reducer devices. %More details on the model and the wind tunnel can be found in \cite{PhdYajun24}. 
%% Coordinate system
A local coordinate system ($x, y, z$), shown in blue in Fig.~\ref{fig:SetUp}, is defined based on a $z$-axis normal to the ground, a $y$-axis normal to the lateral sides of the body, and $x$-axis completing the right-handed triad, with its origin located at the centre of the body’s base. Notably, the $x$-axis corresponds to the direction of the body’s velocity in equivalent road testing with crosswind.

Henceforth, $h$, $u_{\infty}$, and $h/u_{\infty}$ will be used as characteristic length, velocity and time scales respectively. We denote with an asterisk $^*$ the dimensionless variables defined from these characteristic scales.

\subsection{Flaps geometry and mechanics}
\label{subsec:Exp_Confs}

Our study analyses the effect of adaptive rotary flaps attached to the trailing edges of the model (see Fig.~\ref{fig:SetUp}($d$) and how they perform in comparison with classical, rigid systems. To that aim, we test two geometric arrangements: flaps on the top and bottom (TB) edges, or on the left and right (LR) edges. More precisely, we analyze the influence of the degrees of freedom and mechanics of the flap deformation by testing rigid panels attached to the model with either rigid (RF) or elastic hinge (1HF and 2HF) fixations. Depending on the elastic joint design used, see Fig.~\ref{fig:FlapDetailed}, these flaps may  exhibit only bending displacement (1-degree-of-freedom hinged flaps, 1HF) or both bending and torsional displacement (2-degrees-of-freedom hinged flaps, 2HF). To calculate the bending angle, $\theta_b$, and the torsion angle, $\theta_t$, the variable $z_f$ is defined, which represents the position along the $z$-axis of each point that forms the %leading 
trailing edge of the flap. As can be seen in the rear view of Fig.~\ref{fig:SetUp}($f$), the $z_f$ coordinate of the hinged flap’s %leading
trailing edge may vary along the $y$-axis if torsional deformation is present. In this case, the uppermost point of the flap is identified as $\text{max}(z_f^\text{HF})$, the lowest point as $\text{min}(z_f^\text{HF})$, and the average position of the leading edge is denoted by $\overline{z}_f^\text{HF}$. Accordingly, the bending and torsion angles are calculated respectively as follows (see Fig.~\ref{fig:SetUp}($f$) for clarity):
\begin{equation} \label{eq:thetab}
\theta_b=\arcsin\frac{\overline{z}_f^\text{HF}-{z}_f^\text{RF}}{l_f}
\end{equation}
\begin{equation} \label{eq:thetat}
\theta_t=\arcsin\frac{\text{max}(z_f^\text{HF})-\text{min}(z_f^\text{HF})}{w}
\end{equation}
Note that, for the calculation of $\theta_b$, the position $\overline{z}_f^\text{RF}$ corresponding to the fully rigid flap is used. This flap remains flat during the tests, so $\overline{z}_f^\text{RF} = 0.5$.

The 1HF elastic joint is composed of a calibrated-thickness flexible foil that acts as a hinge between the base of the body and a rigid vibrating plate, resulting in a rotational system with one-degree-of-freedom around the $y$-axis, i.e. it can only bend, see Fig.~\ref{fig:FlapDetailed}($a$). The same system has been used in previous work: \citep{CamachoSanchez23} and \citep{Garcia23}. In contrast, the 2HF flexible joint consists of a plate attached to the base of the body %, on which some slots have been applied near the edge of the base, giving rise to the
with a system known as Lamina Emergent Torsional (LET) joint, which has been previously analysed in the literature \citep{Jacobsen2009, Qiu2016, Chau2020}, as shown in Fig.~\ref{fig:FlapDetailed}($b$). The joint pattern allows the flap to rotate around the $y$-axis (bending motion) and the $x$-axis (torsional motion), making it a two-degree-of-freedom system. Additionally, tensile and compressive movement (longitudinal along the $x$-axis) is also possible; however, this is considered negligible due to the virtually non-existent loads experienced by the flap in that direction.
%The 1HF flexible joint consisted of an embedded flexible foil of calibrated thickness, while the 2HF device employed a flexure hinge capable of bending and torsion, referred to as the Triple Lamina Emergent Torsional (LET) system \citep{Qiu2016} (see Fig.~\ref{fig:FlapDetailed}). 
The LET parameters are $a=0.005h, b=0.006h, c=0.07h$ and $ d=0.855h$. The flaps span the full width of the base and have a fixed total length, $l_{f} = 0.5h$, and thickness, $e_{f} = 0.0065h$. 
The flaps are made of PLA through 3D printing (20\% infill and 0.15~mm layer thickness), resulting in a density of $\rho_f \simeq 430~\text{kg/m}^3$, which fixes a mass ratio of $m^{*}=\rho_f / \rho \simeq 360$ for our experiments. Given this large mass ratio, the fluid damping of the motion is very small. Note that the flaps are attached to the Ahmed body using small wedges, which do not alter their mechanical properties and prevent large static deflection of the flaps due to the effect of gravity in some geometrical arrangements.

%\tcg{When a flap is placed horizontally, the gravity produces a positive deflection of about 10$^\circ$. In order to reduce this static deformation, small wedges are inserted between the body base and the flap bracket. This correction brings the static deformation to 4$^\circ$ approximately. These wedges do not alter the mechanical properties of the flaps, just their static position.} %The porous LET structure was observed to have no significant influence on our results.}% 

%\subsubsection{Mechanical characterization of the flaps}
%\label{subsubsec:Exp_MechProp}
Series of free decay tests are conducted to characterize the mechanical properties of the flexibly-hinged flaps. This experimental procedure determines the dynamic characteristics of our system, such as the natural frequency, $f_n$, and the damping coefficient, $\xi$, in response to an initial disturbance as in \cite{CamachoSanchez23, Garcia23}. In this test, the system is first displaced from its equilibrium position by applying an initial deflection. Once the initial disturbance is applied, the system was allowed to oscillate freely without any external forcing, and the subsequent response is recorded over time. The natural frequency is the dominant frequency of that oscillation and %the rate at which the system oscillates %and is a key characteristic of the system's dynamics $f_n=1/2\pi \sqrt{k/m}$, where $k$ is the stiffness of the system and $m$ is the mass. 
the damping coefficient $\xi$ is the decay rate over time of the oscillation amplitude. %over time and was calculated based on this rate of decay. 

\begin{figure}[h]
\centering
\includegraphics[width=0.85\textwidth]{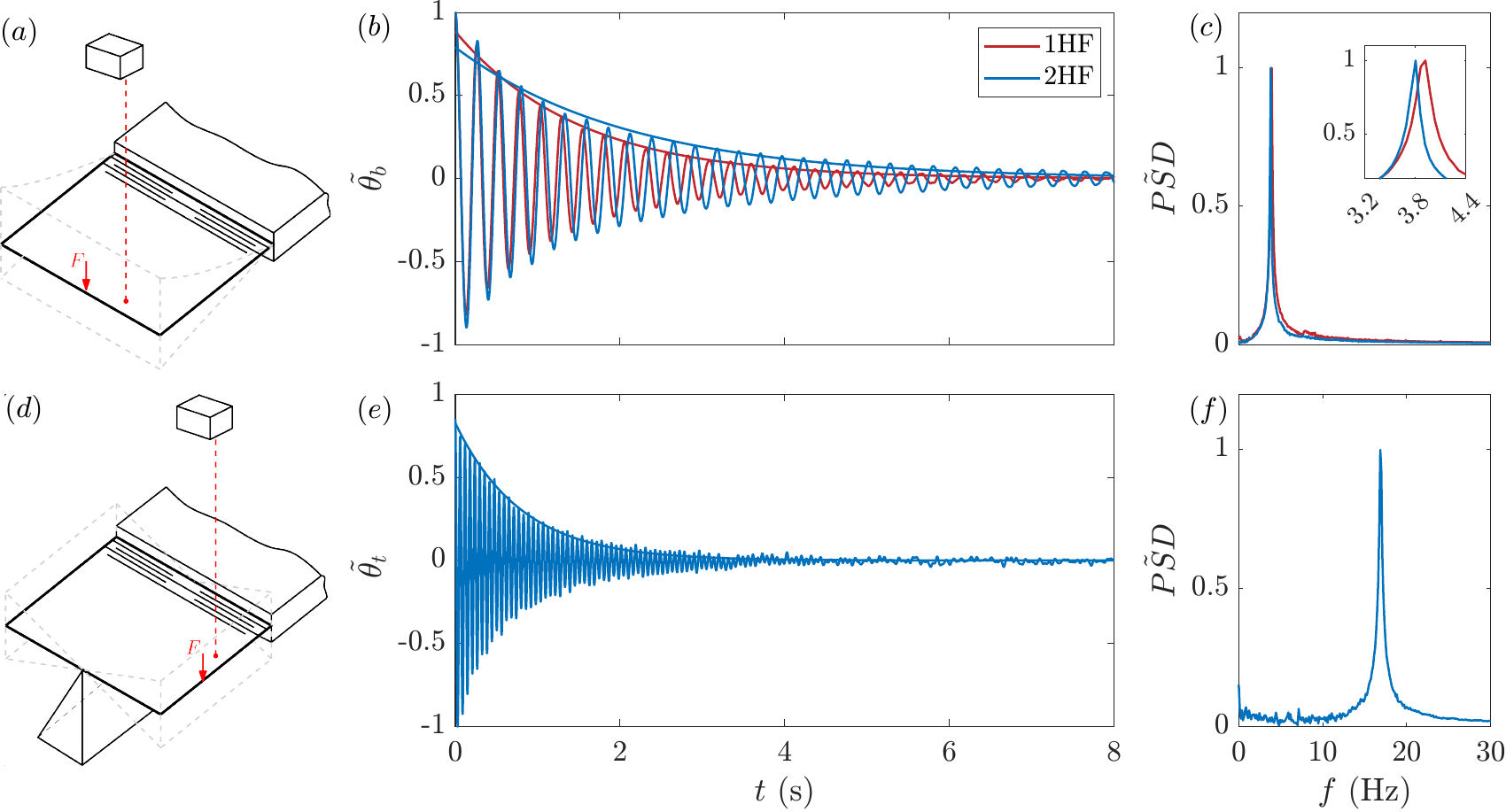}
\caption{\label{fig:FDT} Sketch of the performed free-decay tests for bending ($a$) and torsional ($d$) motions. Temporal evolution and power spectral density (PSD) of the response of the flap in terms of bending ($b, c$) and torsional ($e, f$) angles.}
\end{figure}

Figure~\ref{fig:FDT}($a,d$) illustrates a sketch of the free decay tests performed. Here, the flaps are installed at the base of the body and excited by an initial perturbation, represented by red arrows in Fig.~\ref{fig:FDT}. This induces the free vibration of the system, which is measured using a precision laser distance sensor (model LK-G402 from KEYENCE). In the case of the 2HF device, with two degrees of freedom, it is essential to decouple the two motions, bending and torsion, in order to accurately determine the mechanical parameters of each one. The bending characteristics (natural frequency: $f_{n,b}^{\text{1HF}}$, $f_{n,b}^{\text{2HF}}$ and damping coefficient: $\xi_{b}^{\text{1HF}}$, $\xi_{b}^{\text{2HF}}$) are obtained by the procedure illustrated in Fig.~\ref{fig:FDT}($a$). Conversely, the bending motion of the 2HF is avoided by placing a wedge in the center of the flap along the $x$-axis. This set-up ensures that the flap exhibits only torsional motion, allowing the calculation of the properties associated with that movement (natural frequency: $f_{n,t}^{\text{2HF}}$ and damping coefficient: $\xi_{t}^{\text{2HF}}$). For the 1HF devices, only the free-decay tests corresponding to Fig.~\ref{fig:FDT}($a$) are performed.
%
%First, we obtained the bending natural frequency of the 1HF system ($f_{n,b}^{\text{1HF}}$) as well as its damping coefficient, $\xi_{b}^{\text{1HF}}$. For the 2HF system, two separate tests were performed for obtaining separately the bending parameters ($f_{n,b}^{\text{2HF}}$ and $\xi_{b}^{\text{2HF}}$) and the torsional properties ($f_{n,t}^{\text{2HF}}$ and $\xi_{t}^{\text{2HF}}$).
%

As the excitation force inducing the vibration of the flaps is applied manually, the initial angular displacements, i.e. $\theta_b^\text{1HF}(t=0)$, $\theta_b^\text{2HF}(t=0)$, and $\theta_t^\text{2HF}(t=0)$; vary between tests. However, the obtained mechanical properties are identical regardless of the magnitude of the applied force. To provide the reader with a clear visual comparison of the free-decay test results between the 1HF and 2HF systems, normalised values of the flap deflection by the initial displacement are presented in Fig.~\ref{fig:FDT}. This normalization rescales the differences caused by the applied initial deformations, making it possible to directly compare the response of each system in terms of natural frequency and damping. The temporal evolution of the normalised bending $\tilde{\theta}_b$ and torsional $\tilde{\theta}_t$ angular displacement are shown in Fig.~\ref{fig:FDT}($b,e$) respectively. The response curves exhibit a damped oscillatory motion, where the oscillation amplitudes decay over time due to the damping. The normalised power spectral density ($\text{PSD}$) of the respective signals are presented in Fig.~\ref{fig:FDT}($c,f$), providing the dominant frequencies of the response in each case. 

The free decay bending tests provide very similar results for the 1HF and 2HF, both in natural frequencies ($f_{n,b}^{\text{1HF}} \simeq f_{n,b}^{\text{2HF}}$ and in damping coefficients $\xi_{b}^{\text{1HF}} \simeq  \xi_{b}^{\text{2HF}}$), as evidenced by the results presented in Table \ref{tab:Flaps} and the inset in Fig.~\ref{fig:FDT}($c$). The design and manufacturing of the devices is conceived to produce equivalent systems with similar mechanical properties, to directly compare the role of the degrees of freedom in our problem. Hence, the reduced velocity, $U^*$, defined as the ratio between the free-stream velocity and the averaged velocity of the flexible flap motion, can be computed. In the following, the reduced velocity is defined from the bending motion as $U^{*}_{}=u_{\infty}/(h f_{n,b})$, so that $U^{*}_{}=21.6$.%, and torsion, $U^{*}_{t}=u_{\infty}/f_{n,t}h$ motions.

\begin{table}[t]
\centering
\begin{tabular}{ccc}
\hline
                   & 1HF & 2HF \\ \hline
                        $f_{n,b}$ (Hz) &   3.70  &  3.76  \\
                        $\xi_{b}$ &   0.025  &  0.019  \\
                        $f_{n,t}$ (Hz) &   ---  &  16.85  \\
                        $\xi_{t}$ &   ---  &  0.009  \\
                        \hline

%\multirow{-4}{*}{0.02} & V & \cellcolor[HTML]{FFFFFF}-0.0008 & \cellcolor[HTML]{FFFFFF}-0.0428 & \cellcolor[HTML]{FFFFFF}-0.0420 & 0.1756 \\ \hline
\end{tabular}
\caption{Mechanical parameters of the flaps obtained from the free-decay tests.}
\label{tab:Flaps}
\end{table}

\subsection{Measurements}
\label{subsec:Exp_Meas}

\subsubsection{Force coefficients}
\label{subsubsec:Exp_Forces}

%% Forces
The aerodynamic force exerted on the model is measured by a six-component force balance (F/T Sensor: ATI Gamma IP65), as shown in Fig.~\ref{fig:SetUp}. The force balance is fixed to the rotation stage, and measures the aerodynamic loads of the body in three directions $\text{\textflorin}_x$, $\text{\textflorin}_y$ and $\text{\textflorin}_z$ defined by the coordinate system displayed in Fig.~\ref{fig:SetUp}($a-c$). The sampling frequency is 1 kHz per component. The model frontal area, $hw$, is used to calculate the force coefficients: 
\begin{equation} \label{eq:forces}
c_i=\frac{2~\text{\textflorin}_i}{\rho u_\infty^2 hw},~~ i=x,y,z.
\end{equation}
The manufacturer resolution is 0.025 N for $\text{\textflorin}_x$, $\text{\textflorin}_y$, and 0.05~N for $\text{\textflorin}_z$. Thus, at the usual flow velocity $u_{\infty}$ = 16 m/s, these resolutions translate to a precision around $3 \cdot 10^{-3}$ for $c_x$, $c_y$ and $6 \cdot 10^{-3}$ for $c_z$. However, by using calibrated weights, the accuracy on time-averaged measurements was actually as high as 0.001 in term of drag coefficient, as reported in \cite{Fan2022} using the same experimental set-up. The natural frequency of the system composed of the model and the force balance (measured from the force balance signal) is $f_n=9$~Hz, which in non dimensional units corresponds to $f_n^*=f_{n}h/u_{\infty}=0.11$. No attempts have been made to extract aerodynamic force fluctuation with such a low frequency response of the force balance, and only mean force will be analysed. %Our study is focused on the effect of the different tested flaps in terms of drag reduction at different yaw angles, then, the aerodynamic coefficients will be defined in the direction of the model. %\tcr{What is the point of introducing a Strouhal number ? We already defined $f^*$ as a non dimensional frequency. The Strouhal number is also a non dimensional frequency but associated with periodic shedding. I would suggest to keep $f^*$, which is clear and avoids (discussable) additional definitions in the paper.}

\subsubsection{Base pressure distribution}
\label{subsubsec:Exp_pressure}
Pressure is measured using taps connected via vinyl tubing to a pressure scanner (Scanivalve ZOC33/ 64PX). The 20 pressure taps are equally spaced at the base with distances $\delta_y$ = 53.3~mm and $\delta_z$ = 45~mm as shown in Fig.~\ref{fig:SetUp}($c$). Vinyl tubes connecting taps and scanner never exceeded 50~cm, acting as a natural low-pass filter with a cut-off frequency of approximately 50 Hz ($f_\text{cut-off}^* \simeq 0.625$). The sampling frequency was 1 kHz per channel. The static pressure of the test section, $p_\infty$, was recorded from a pair of Pitot tubes placed at the inlet of the test section, and connected to a Precision Manometer (FCO560) to monitor the dynamic and static pressure for calculating the free-stream velocity, $u_\infty$. Thus, the instantaneous pressure coefficient is obtained as
\begin{equation} \label{eq:cp}
c_{pj}=\frac{2(p_j-p_\infty)}{\rho u_\infty^2},
\end{equation}
and the base suction (or base drag) coefficient \citep{Roshko93} is estimated by means of 
\begin{equation} \label{eq:cb}
c_B=-\frac{1}{n}\sum_{j=1}^n c_{pj}
\end{equation}
being $n=20$ the total number of base pressure taps. 
The pressure scanner accuracy is reported to be $\pm3.75$ Pa by the manufacturer as $0.15\%$ of the full scale 2.5 kPa. To improve this value, the pressure scanner is periodically calibrated in a range of $\pm$ 200~Pa. The calibration is made using a Precision Manometer Calibrator (FCO560) having an accuracy of 0.01~Pa. The pressure scanner accuracy is estimated from the measurement of 40 Pa delivered by the calibrator, that is the typical value of the base drag in this present study. All the time-averaged base pressure values, fall within the range 40$\pm$0.5~Pa. We thus estimate the pressure coefficients accuracy to be $\pm 0.005$.

Additionally, following previous studies \citep[see e.g.][]{Grandemange13b,Lorite20b} the wake asymmetry can be quantified with help of the components $g_y$ and $g_z$ of the base pressure gradients calculated as
\begin{equation} \label{eq:gy}
g_y= \frac{1}{2}h\left[\frac{(c_{pB}-c_{pA})+(c_{pD}-c_{pC})}{y_{B}-y_{A}}\right],
%g_y=h\frac{\partial c_p}{\partial y} \simeq \frac{1}{2}h\left[\frac{(c_{p,15}-c_{p,14})+(c_{p,7}-c_{p,6})}{y_{15}-y_{14}}\right],
\end{equation}
\begin{equation} \label{eq:gz}
g_z= \frac{1}{2}h\left[\frac{(c_{pA}-c_{pC})+(c_{pB}-c_{pD})}{z_{A}-z_C}\right].
%g_z=h\frac{\partial c_p}{\partial z} \simeq \frac{1}{2}h\left[\frac{(c_{p,14}-c_{p,6})+(c_{p,15}-c_{p,7})}{z_{15}-z_7}\right].
\end{equation}
Note that the four pressure measurements at $A, B, C, D$ are sufficient to estimate accurately the base pressure gradients since the wall pressure distribution in the separated area is at first order almost constant in one direction and affine in the other perpendicular direction (direction of the asymmetry) \citep[see e.g.][]{Barros2017, Fan2022}.

Force and base pressure measurements are recorded during 20~s to ensure accuracy better than 0.5$\%$ for mean values. To avoid drift effects especially from the force balance measurements, periodically updated references are made following the procedure detailed in \cite{Fan2022}. All coefficients and gradients are low-pass filtered with a moving window of $t_w^*=3.95$, implying that the dynamics is resolved at low frequencies such that $f^*<\frac{1}{t_w^*} \approx 0.25$.

\subsubsection{Wake velocity field}
\label{subsubsec:Exp_StereoPIV}
Stereo-Particle Image Velocimetry (2D-3C PIV) is employed to measure 3 components (3C) velocity fields $\textbf{u}=(u_{x}, u_{y}, u_{z})$ in a 2D plane in the wake behind the model. A high-performance double pulse laser model Litron LPY704-100 PIV (100 mJ/pulse, 100 Hz, 532 nm) is used to generate a laser sheet in the test section. To precisely control the laser sheet's position and thickness ($\simeq 1$mm) within the measurement volume, a $90^{\circ}$ mirror, a cylindrical lens (-10 mm) and a collimator are used.  The vertical laser sheet is placed as a plane $x^*= 0.915$ downstream the base. The laser is synchronised through a LaVision{\textregistered} PTU-X with a pair of high-speed cameras Phantom VEO 340L with a resolution of 1600 × 2560 pixels, equipped with a focal lens objective of 50 mm, f/5.6 and two Scheimpflug mounts. %In addition,  a polariser filter is attached over the lenses to avoid the laser reflections. 
The Stereo-PIV calibration is conducted using a 3D checkerboard calibrated plate, provided by LaVision{\textregistered}, to properly set the camera focus, Scheimpflug adjustment and the laser plane position and width, obtaining a scale of 4.62 px/mm. Moreover, the incoming wind is seeded using an oil-droplet generator, with tracers of diameter $\simeq 2 \mu$m. The laser sheet is pulsed with time delays of $dt = 10\mu$s, and the set-up records 600 pairs of images at 15 Hz, ensuring an adequate number of images for obtaining time-averaged velocity fields. After correctly setting the region of interest and performing simple image preprocessing, velocity vectors are obtained from multi-pass PIV correlation in interrogation windows sized 64 $\times$ 64 pixels with an overlap of 75$\%$. The resulting velocity vectors are spatially distributed in a grid of $103 \times 219$ points,  with a resolution around $1.7\%$ of the body’s height.% for both measurement planes. 

To ensure the reliability and accuracy of our results, three independent PIV tests are conducted for each experiment, yielding 51 tests in total. Finally, we checked our velocity measurements correlation-based uncertainty \citep{Wieneke2015}, obtaining errors below 0.7\% of the free-stream velocity and below 2.5\% for the other velocity components and slow velocity regions, which warrants the robustness of our PIV measurements.

\subsubsection{Flap deformation}
\label{subsubsec:EXP_Flap}
The same Stereo-PIV principle is used to reconstruct the deformation of the flaps. To do that, we painted between 10 and 15 white fiducial dots at the trailing edges of the flaps to record their positions. The experimental procedure consists first, to yaw the Ahmed body at a given angle $\beta$, second, to place the calibration plate parallel to the body base at the trailing edge of the flaps and third, to adjust the cameras. % Next, the camera focus, position and orientation is adjusted as well as the Scheimpflug mounts. 
%The calibration is performed using the plate reaching errors below 0.2 px.
Due to negligible streamwise motion of the flaps, the trailing edges marked with the white dots remain in the calibration plane during the wind testing. We can thus reconstruct the flap motion in the laboratory frame.
%to track the deformation of the flap with this calibration. 
Before changing the yaw, all the tested flaps are also recorded at lower flow velocities. The measurements are acquired at 100~Hz during 20~s which corresponds to 80 cycles at the bending natural frequency of the flaps. %These measurements were made to estimate the flaps averaged angular position and the corresponding amplitude of angular vibrations. These angles were obtained with respect to the position of the trailing edge of the rigid flaps for each yaw condition.

The dynamic response of the flaps is analysed through additional measurements using laser displacement sensors (Keyence model LK-G402) with a linearity of 0.05~mm and a sampling frequency of 1000 Hz (see Fig.~\ref{fig:FlapDetailed}). To avoid interfering with the flow and introducing disturbances, the laser sensors must be placed outside the wind tunnel which is possible only from the top of the test section which has optical access. Thus, only the motion of the top flap is investigated. 
%As the test section floor do not have optical access and the bottom flap was not measured with this technique. 
% The sensors measure the linear displacement of the flap at two locations (one centered and one off-centered), from which, after adequate trigonometric transformations, the instantaneous angular deflections are obtained.
The linear resolution provided by the laser sensor translates into an angular accuracy of $0.03^\circ$.%0.0005~rad.

\section{Results}
\label{sec:Results}
As a reminder, $h$, $u_{\infty}$, and $h/u_{\infty}$ are used as characteristic length, velocity and time scales respectively. We denote with an asterisk $^*$ the dimensionless variables defined from these characteristic scales. Moreover, instantaneous variables are denoted by a lowercase letter $v$, their time-averaged values by a uppercase letter $V = \overline{v}$, and their fluctuating part by $v'=v - V$. The spatial average is denoted by $\langle v \rangle$. %\tcr{The modulus of a given variable will be expressed as $|v|=\sqrt{v^2}$}. 
All vector variables, whether instantaneous ($\Vec{v}=\bm{v} $) or time-averaged ($\Vec{V}=\bm{V}$),  are represented in bold font and their components specified with subscripts $x,y,z$.

The baseline is defined as the model without any flaps as shown in Fig.~\ref{fig:SetUp}($d$, left). The baseline aerodynamic properties have already been reported in \cite{Fan2022} and \cite{PhdYajun24} where a wide range of body attitudes (pitch, yaw and ground clearance) have been explored. For the purpose of the present study, baseline measurements have been repeated in \S \ref{sec:baseline} for the specific attitudes investigated with the flaps and will be used in the following sections as a reference to study the aerodynamic effects of the flap arrangements (\S \ref{sec:arrangements}) and the degrees of freedom of the flap motion (\S \ref{sec:DoFs}).

\subsection{Baseline case}
\label{sec:baseline}

\begin{figure}[t]
\centering
\includegraphics[width=0.6\textwidth]{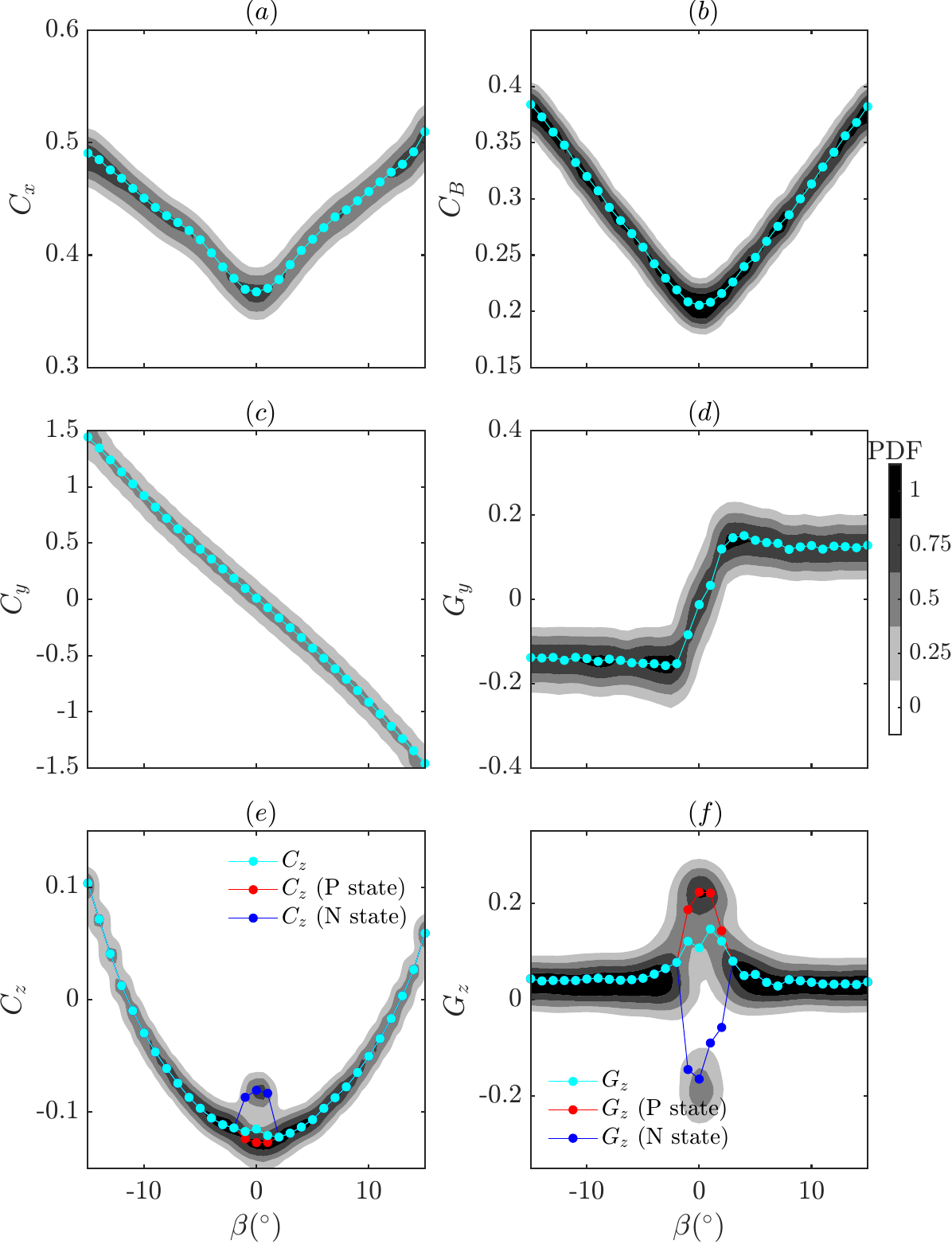}
\caption{\label{fig:BaseAero}$(a-f)$ Evolution of the time-averaged (cyan symbols) and Probability Density Function (PDF) contours of the force and pressure coefficients with the yaw angle $\beta$ for the baseline case (B) at $Re= 2.13 \cdot 10^5$. Force and pressure coefficients: Drag, $c_x$, base drag, $c_B$, lateral force coefficient, $c_y$, horizontal base pressure gradient, $g_y$, vertical force coefficient, $c_z$,  and vertical base pressure gradient, $g_z$. Blue and red points represent the conditionally averaged values for N and P state respectively in $z$-axis.}
\end{figure}

\begin{figure}[t]
\centering
\includegraphics[width=1\textwidth]{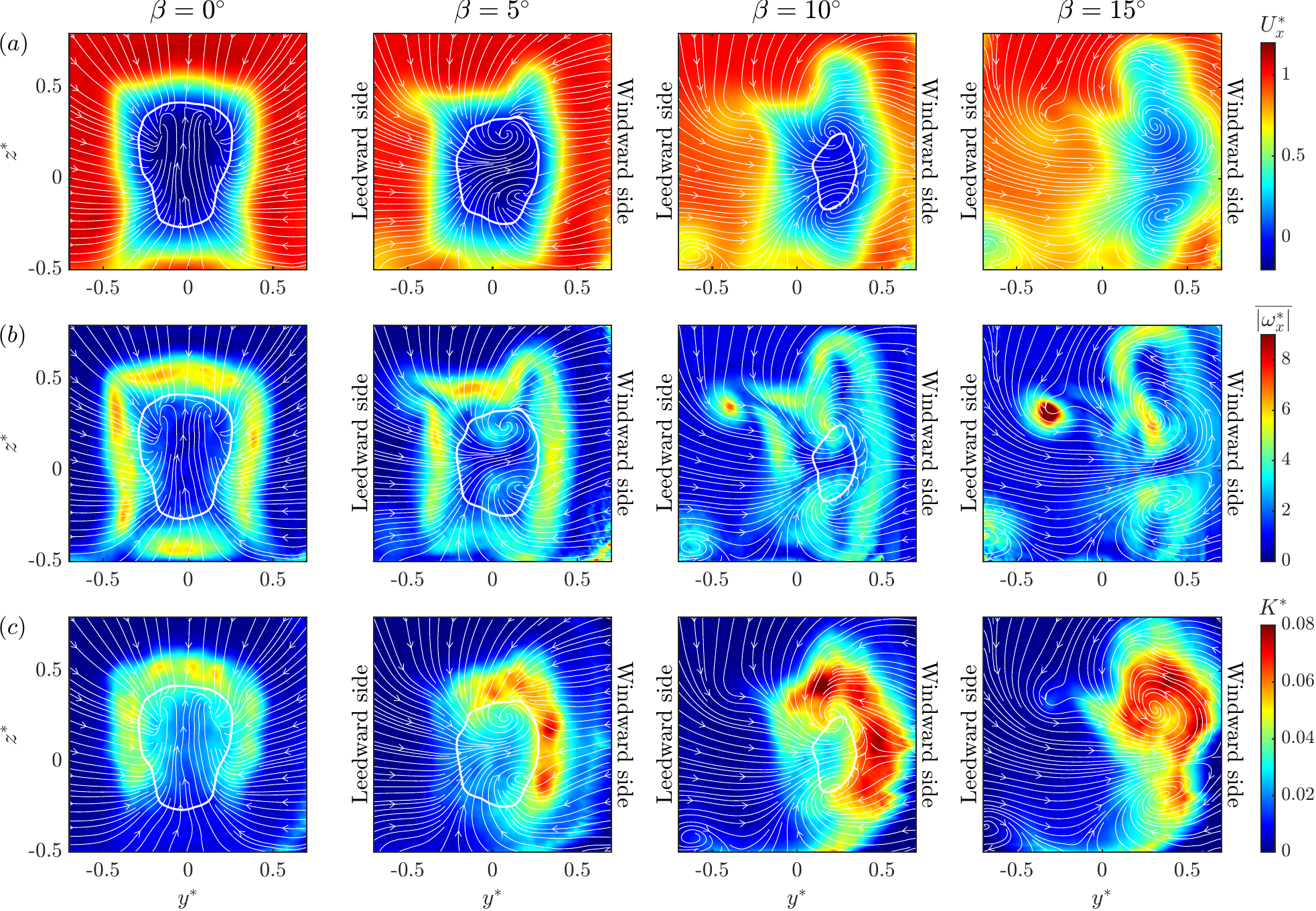}
\caption{\label{fig:BasePIV} For $\beta = [0:5:15]^\circ$ at $Re= 2.13 \cdot 10^5$: Time-averaged contours of $(a)$ streamwise velocity, $U_x^*$, $(b)$ magnitude of streamwise vorticity, $\overline{|\omega_x^*|}$ , and $(c)$ turbulent kinetic energy, $K^*$, at $x^{*}$ = 0.915 for B configuration. Thin white lines illustrate the flow streamlines ($U^{*}_{y}$, $U^{*}_{z}$), while thick white line shows the isoline of $U_x^*=0$.}
\end{figure}

The baseline force and pressure coefficients are shown in Fig.~\ref{fig:BaseAero} as stacked probability density functions (PDFs) versus yaw value. This representation captures eventual presence of bistable behaviour of the Reflectional Symmetry Breaking (RSB) mode which is ubiquitous in the wake of three-dimensional bluff bodies with a rectangular base \citep{Grandemange2012}. In addition, plots include the averaged value of the coefficient with cyan symbols to provide the mean trend with the yaw. In the flow conditions where bi-stability is observed, averaging conditioned by the vertical pressure component $g_z$ is performed to distinguish the mean coefficients of the P state with positive component (red symbols) and the N state with the negative component (blue symbols).

The drag coefficient in Fig.~\ref{fig:BaseAero}($a$) shows the classical trend of drag increase with yaw reported for squareback bodies \citep{Hassaan18, Mcarthur2018} but also more generally for any real car geometries \citep{Howell_2015}. The drag increase is clearly associated with the increase in base suction as shown in Fig.~\ref{fig:BaseAero}($b$) : the larger the yaw, the larger the base suction. This effect is known to be related to three dimensional separations along the body edges between its four lateral sides which produces intense longitudinal vortices lowering the base pressure or identically, increasing the base suction.

Both lateral force components $C_y$ and $C_z$ are mainly due to the pressure loads of the four body sides, however a wake effect can be identified with the base pressure gradient \citep{Fan2022}. The side force in Fig.~\ref{fig:BaseAero}($c$) is seen to be at first order a linear function of the yaw, however, at second order
a wake effect evaluated through the horizontal base pressure gradient component showing an abrupt change in Fig.~\ref{fig:BaseAero}($d$) within the small yaw range of $\pm 3^\circ$ is responsible for the hyperbolic behaviour of $C_y~(\beta\approx 0^\circ)$. The lift force, of a quadratic form in Fig.~\ref{fig:BaseAero}($e$) presents a clear bistable behaviour in the same yaw range of $\pm 3^\circ$. This is again due to a wake effect as revealed by the vertical base pressure gradient in Fig.~\ref{fig:BaseAero}($f$). In \cite{Fan2022}, the wake transition that operates at $\beta \approx \pm 3^\circ$ is a global rotation of the mean recirculating region by an angle close to $\pi/2$, where the strong asymmetry in the vertical direction observed at small yaw tilts towards the horizontal direction at larger yaw. For the small yaw attitudes, the P state is dominant during the bistable dynamics resulting in mean properties closer to the P state than to the N state.

The near wake structure of the mean flow at $x^*=0.915$ is shown in Fig.~\ref{fig:BasePIV} for different yaws. The strong velocity gradients represented in green in Fig.~\ref{fig:BasePIV}($a$) is representative of the surface separation that delimits the recirculating area, that appears in the blue colour range, to the external flow belonging to the red colour range. The back-flow (where $U_x^*<0$) is surrounded with the white continuous line, corresponding to the isoline $U^*=0$. These velocity fields confirm the $\pi/2$ global rotation of the cross flow inside the recirculating area between the two attitudes $\beta=0^\circ$ and $\beta=5^\circ$. In addition, the surface separation gets significantly deformed at its corners as yaw is increased. The basic mechanism reported in \cite{Mcarthur2018, Hassaan18, Booysen22} involves the three-dimensional separations along the edges between the four lateral body sides that are intensified as the yaw is increased. The mean modulus of the streamwise vorticity component $\overline{|\omega^*_x|}$ in Fig.~\ref{fig:BasePIV}($b$) indicates a strong longitudinal vortex on the top leeward side of the base at $\beta=10^\circ$ that is greatly intensified at $\beta=15^\circ$. Such a structure contributes to lower the pressure in the base region or equivalently to increase the base suction coefficient as seen in Fig.~\ref{fig:BaseAero}($b$).

Eventually, the local mean fluctuation measured as the mean turbulent kinetic energy $$K^{*} = \sqrt{\overline{{u'^*}^2}+\overline{{v'^*}^2}+\overline{{w'^*}^2}}$$ is shown in Fig.~\ref{fig:BasePIV}($c$). As yaw is increased, the fluctuation intensifies and concentrates towards the windward side.

\subsection{TB vs. LR flaps arrangement}
\label{sec:arrangements}
\begin{figure}[t]
\centering
\includegraphics[width=0.75\textwidth]{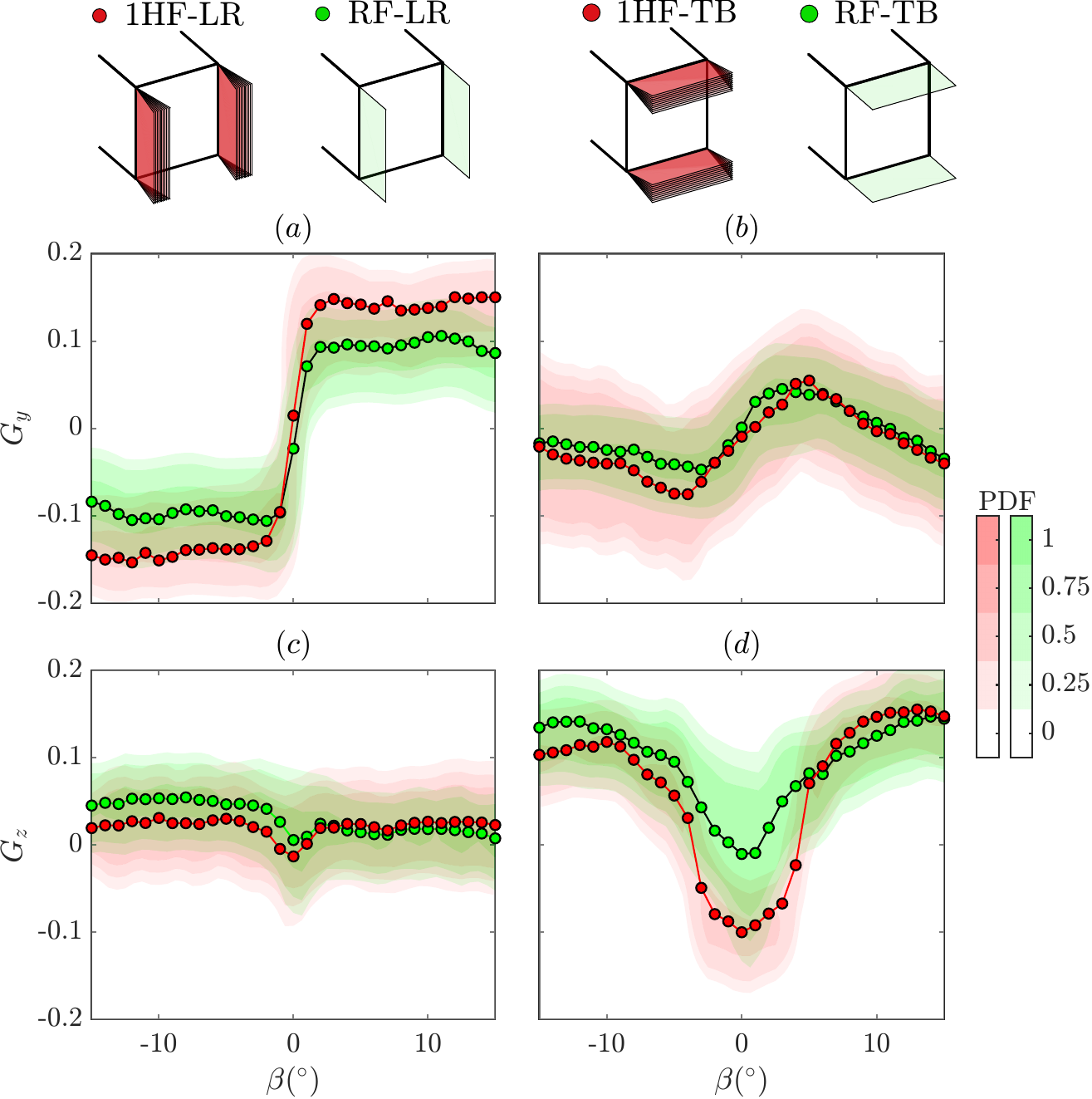}
\caption{\label{fig:Comp_LRTB_GYGZ}Time-averaged (symbols) and Probability Density Function (PDF) contours of the base pressure gradient components ($a,b$) $g_y$ and ($c,d$) $g_z$ vs. yaw for the LR arrangement ($a,c$) and TB arrangement ($b,d$) at $Re= 2.13 \cdot 10^5$. For both arrangements, rigid flaps correspond to green colour and elastic hinged flaps to red colour.}
\end{figure}
\begin{figure}[t]
\centering
\includegraphics[width=0.75\textwidth]{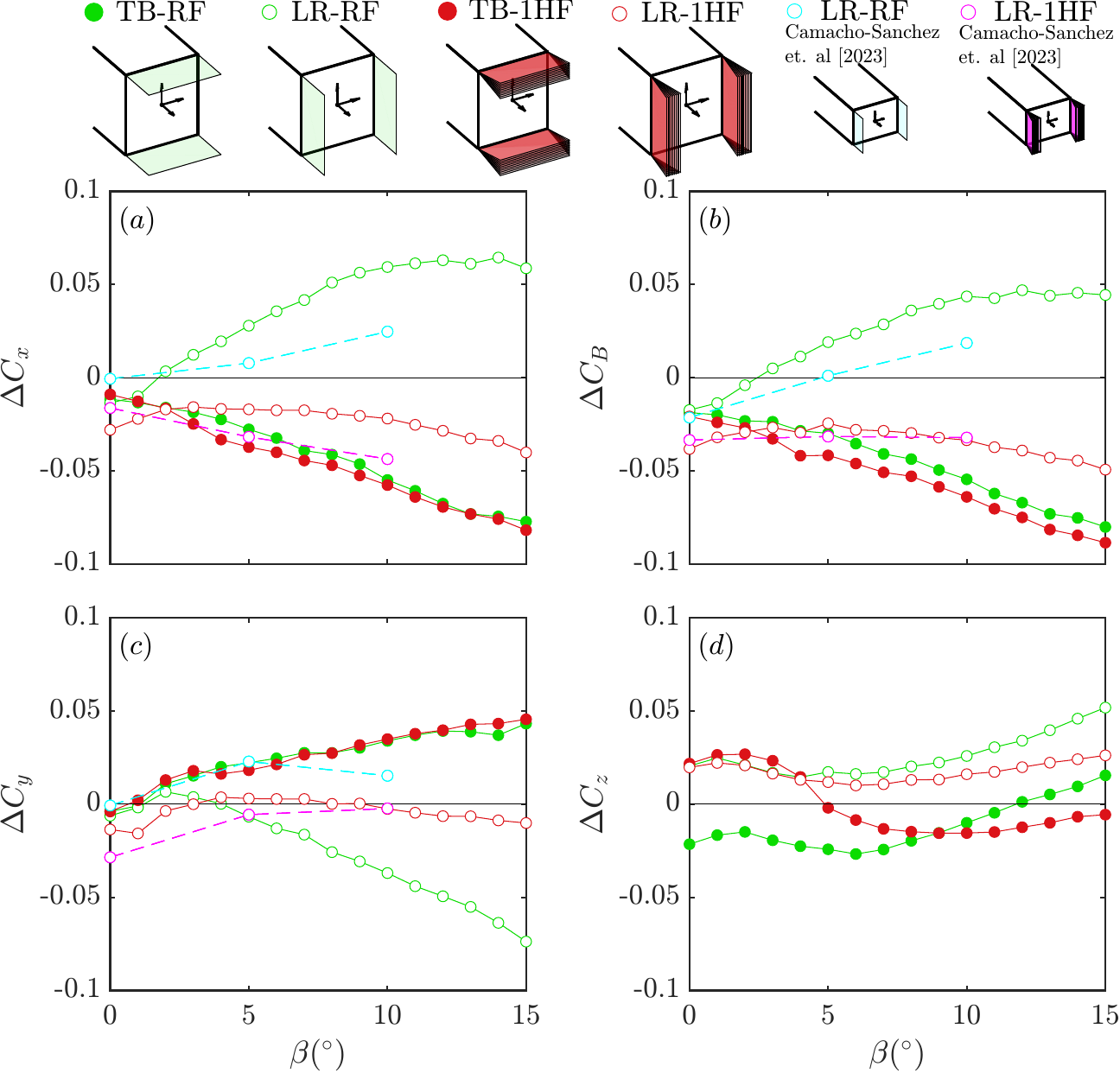}
\caption{\label{fig:Comp_LRTB} Evolution of the averaged relative force coefficients ($a,c,d$): $\Delta C_x, \Delta C_y, \Delta C_z$ respectively and base drag ($b$): $\Delta C_B$ with the yaw angle ($\beta$) for the Left-Right (LR) and Top-Bottom (TB) arrangements at $Re= 2.13 \cdot 10^5$. Data from \cite{CamachoSanchez23} are added in ($a, b, c$).}
\end{figure}
The influence of the flaps arrangement, either placed along the vertical edges of the base (LR arrangement) or along the horizontal edges (TB arrangement as depicted in Fig.~\ref{fig:SetUp}($d$) is now investigated. We first compare their effect on the base pressure gradient components in Fig.~\ref{fig:Comp_LRTB_GYGZ}($a,c$) for the LR arrangement and in Fig.~\ref{fig:Comp_LRTB_GYGZ}($b,d$) for the TB arrangement. We can see in Fig.~\ref{fig:Comp_LRTB_GYGZ}($c$) that the LR arrangement suppresses the bistable behaviour of the vertical gradient component. Indeed, the two most probable values at approximately $\pm 0.2$ that was observed in the small yaw range $\pm 3^\circ$ of the baseline in Fig.~\ref{fig:BaseAero}($f$) have been replaced by a single most probable value. As a result, the base gradient vector is almost zero for zero yaw leading to a symmetric pressure distribution meaning a possible suppression of the RSB mode. This situation would be very similar to the known stabilisation using a base cavity \citep{Evrard16}. The only asymmetry with the LR arrangement is produced in the horizontal direction when $\beta \neq 0^\circ$ as shown in Figs.~\ref{fig:Comp_LRTB_GYGZ}($a$). It is likely that the LR arrangement is associated with an asymmetry towards the horizontal direction, or equivalently to a direction perpendicular to the flaps. A similar trend is observed for the TB arrangements, where the component of the pressure gradient in the horizontal direction in Figs.~\ref{fig:Comp_LRTB_GYGZ}($b$) is significantly reduced compared to the baseline (see Fig.~\ref{fig:BaseAero}($d$)), thus increasing the asymmetry towards the vertical direction, again perpendicular to the flaps. Whatever the arrangement, whether in the rigid or elastic flap configuration, the bistable behaviour is clearly suppressed, either due to a stabilisation or a mode selection. It is also observed that due to their adaptive shape, the elastic flaps can produce slightly larger gradient components than the rigid flaps.

The corresponding aerodynamic coefficients of drag, side force, lift force and base drag are shown in  Fig.~\ref{fig:Comp_LRTB}. They are actually represented as variation from the baseline defined as  $\Delta C_i = C_i - C_i ^\text{B}$ where $i=B,x,y,z,$ and $C_i ^\text{B}$ are the aerodynamic coefficients shown in Fig.~\ref{fig:BaseAero}($a-d)$. It can be seen in Figs.~\ref{fig:Comp_LRTB}($a,b$) that all configurations reduces drag and base suction at $\beta=0^\circ$ as all variations are negative. When yaw is increase, very different trends are observed depending on the configuration. First, the rigid LR arrangement shown with empty green symbols produces a drag increase in Fig.~\ref{fig:Comp_LRTB}($a$) which  cannot be retained as a drag-reducing device. In the same arrangement, but with elastic hinges (LR-1HF) displayed as empty red symbols, an almost constant drag reduction of about 0.025 is obtained across the low values of the tested yaw angles, increasing with $\beta>10^\circ$ to reach reductions up to 0.04. This same trend between rigid and bending flaps in the LR arrangement was reported in \cite{CamachoSanchez23} for an Ahmed body with a different aspect ratio ($w/h=1.35$) of which their data points are also included in Fig.~\ref{fig:Comp_LRTB}. While the LR arrangements (empty symbols) display, in general, better aerodynamic performance  
at small yaw (for approximately $\beta<3^\circ$) likely to be due to the suppression of the RSB mode as in \cite{Evrard16}, as discussed above, the TB arrangements (filled symbols) definitely represent a better solution at yaw, with a linear variation of drag reduction and where elastic flaps (filled red symbols) present a small improvement compared to the rigid ones (green filled symbols).

Moving now to the side force effect presented in Fig.~\ref{fig:Comp_LRTB}($c$), we can see that the TB arrangement (filled symbols), whether rigid or elastic, produce the same side force increase with yaw compared to the baseline. Hence, the elastic deformation of the TB arrangement has no effect on the side force. The baseline has a negative side force at positive yaw as shown in Fig.~\ref{fig:BaseAero}($c$), and it is the reduction of the horizontal base pressure gradient component observed in Fig.~\ref{fig:Comp_LRTB_GYGZ}($b$) that explains the increase (or less negative) side force. In others words, the TB arrangements reduce the magnitude of the side force with yaw through a wake effect. In contrast, the side force is very sensitive to the elastic deformation of the LR arrangement as can be seen by comparing the green and red empty symbols in Fig.~\ref{fig:Comp_LRTB}($c$). As expected, the side force is continuously decreased (or more negative) with the lateral side extension of the body produced by the LR rigid flaps (green empty symbols). On the other hand, the adaptive mechanism of the elastic flaps \citep[see][]{CamachoSanchez23} tends to maintain a side force very close to that of the baseline as variations (red empty symbols) are small. Note that the quasi-steady adaption of flexible flaps has been shown to decrease the size and bluffness of the recirculating region, leading to a weakening of the backflow and base pressure recovery (see e.g. \cite{MunozHervas2024}).%\tcr{(Was this explained in the previous paper ?)}

For the lift force in Fig.~\ref{fig:Comp_LRTB}($d$), the LR arrangement (empty symbols) produces an increase on the whole yaw range. This is likely due to a wake effect associated with the small reduction of the vertical base pressure gradient component between the baseline (Fig.~\ref{fig:BaseAero}($f$) and the additional LR flaps (Fig.~\ref{fig:Comp_LRTB_GYGZ}($c$)). %It can be noticed, to consolidate this possibility, that the larger reduction of the vertical base pressure gradient component for the rigid flaps than for the elastic ones  for positive yaws in Fig.~\ref{fig:Comp_LRTB_GYGZ}$c$ is in agreement with a larger lift increase of the rigid flap (empty green symbols) than for the elastic flaps (empty red symbols) in Fig.~\ref{fig:Comp_LRTB}($d$). Thus, the lift change with the LR arrangements is consistent with a wake effect. 
In contrast, the lift change with the TB arrangements is more likely a consequence of the force exerted on the flaps. With rigid TB flaps (green filled symbols), the lift is first reduced for yaws $\beta<12^\circ$ and then increased. This is probably an effect of the lower flap that is at proximity of the floor. With the elastic TB flaps (red empty symbols), the lift behaves completely differently because of the flap deflections are acting as additional lifting surfaces. This will be further discussed in the next section.

\subsection{Increasing the degrees of freedom for the TB arrangement}\label{sec:DoFs}
\begin{figure}[t]
\centering
\includegraphics[width=0.85\textwidth]{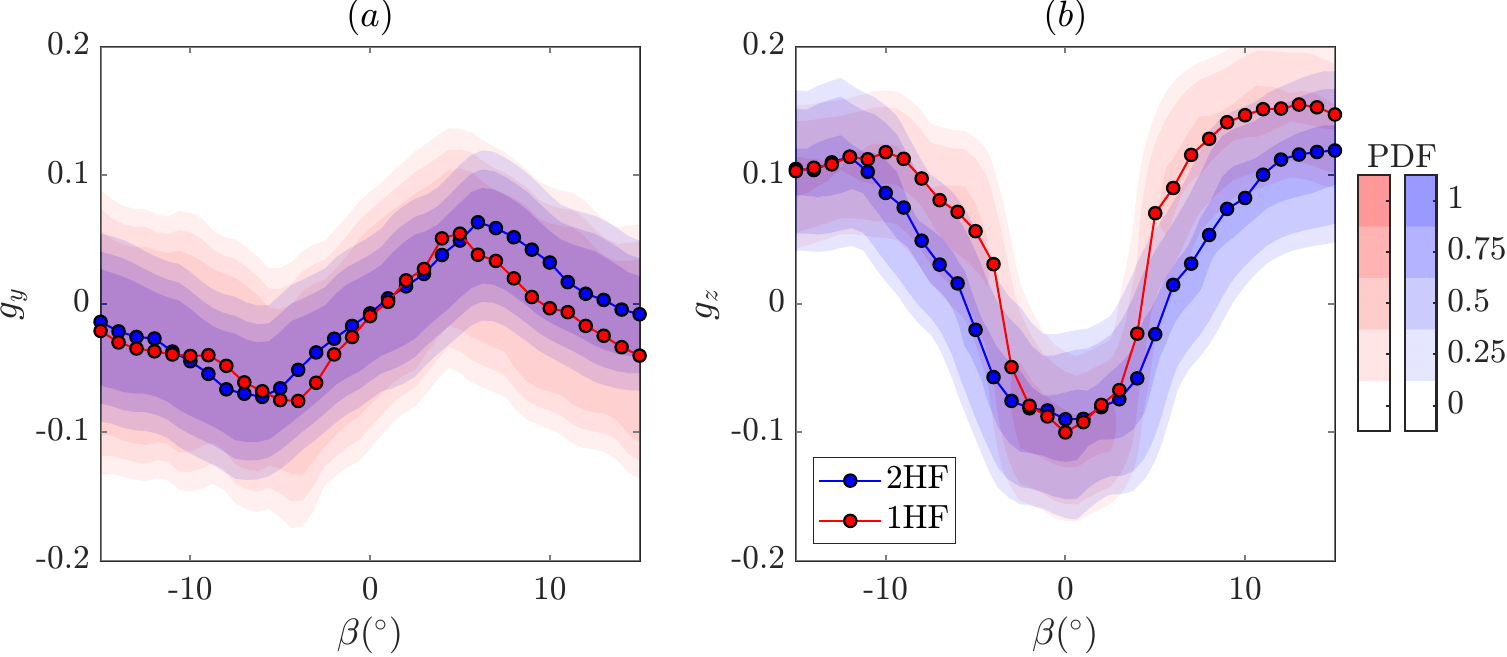}
\caption{\label{fig:Comp_HF_GYGZ} TB elastic flaps arrangement with torsion and bending (2HF, blue colour) compared to bending only (1HF, red colour). Time-averaged (symbols) and Probability Density Function (PDF) contours of the base pressure gradient components, ($a$) $g_y$ and ($b$) $g_z$ vs. yaw at $Re= 2.13 \cdot 10^5$.}
\end{figure}

\begin{figure}[t]
\centering
\includegraphics[width=0.75\textwidth]{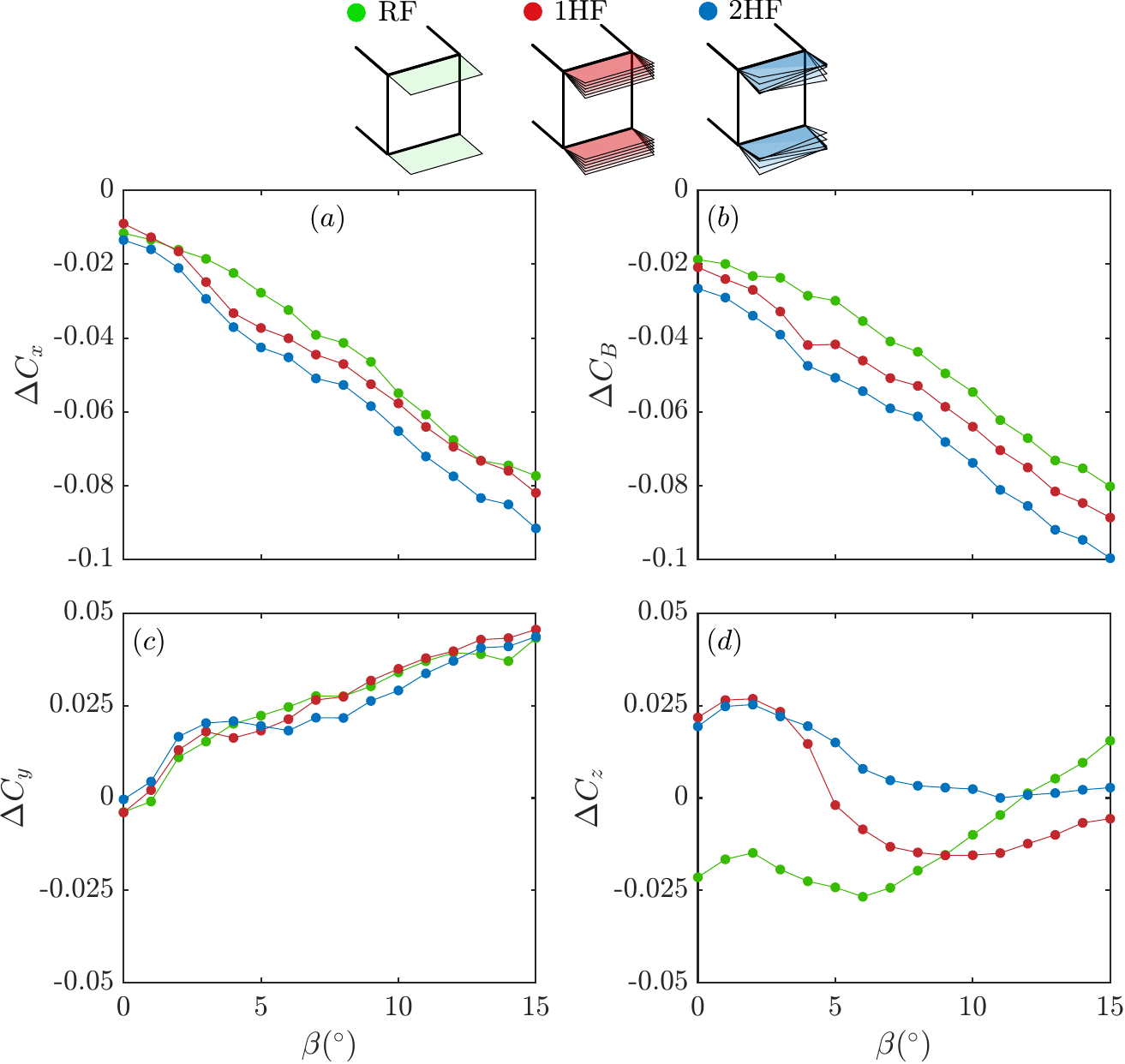}
\caption{\label{fig:Comp_TB} Evolution of the averaged relative base drag ($a$) $\Delta C_B$ and force coefficients  ($b,c,d$) $\Delta C_x, \Delta C_y, \Delta C_z$  for Top-Bottom (TB) arrangement for the cases of Rigid Flaps (RF), bending flaps (1HF) and bending/torsion flaps (2HF) at $Re= 2.13 \cdot 10^5$.}
\end{figure}\label{sec:aero_LRTB}
The TB arrangement has been shown to be the most efficient configuration to reduce drag in cross-flow for yaws $|\beta|>3^\circ$. Therefore, we further study this arrangement by improving its ability to adapt to the cross flow by adding a second degree of freedom at the junction between the flap and the body. It consists of the torsion as depicted in Fig.~\ref{fig:SetUp} (under the 2HF configuration) and Fig.~\ref{fig:FlapDetailed}($b$) with properties given in Fig.~\ref{fig:FDT}($d-f$).

Figure~\ref{fig:Comp_HF_GYGZ} shows the effect of the 2HF configuration compared to the 1HF on the base pressure gradient components. There are no drastic changes, the horizontal components are very similar in Fig.~\ref{fig:Comp_HF_GYGZ}($a$), and the only noticeable difference is for the vertical components where the additional torsion produces a decrease for intermediate yaws (maximum 40$\%$). In the same way as observed with the RF and 1HF systems using the TB configuration, the 2HF flaps also mitigate the bistable behaviour of the wake, fixing a N state ($g_z<0$) at small yaw angles.

Figure~\ref{fig:Comp_TB} shows the variations of the aerodynamic coefficients as a function of the yaw angle $\beta$. In this figure, we have reproduced the RF and 1HF configuration data of the TB arrangement already presented and commented in the previous part. Although, they won't be discussed again, the set of figures has the advantage to show at a glance the effect of increased degrees of freedom from the rigid (RF), bending only (1HF) and bending/torsion (2HF) flaps for the given TB arrangement. A clear improvement of the drag reduction can be seen for the 2HF configuration in Fig.~\ref{fig:Comp_TB}($a$) on the whole range of yaw. Note that, at large yaw, the decrease in the drag becomes progressively more apparent for 2HF configuration, whereas RF and 1HF trends converge. The drag trend is a consequence of the base suction reduction observed in Fig.~\ref{fig:Comp_TB}($b$). Notice that the base suction reduction is always larger than the drag reduction. The reason is the additional drag produced by the flaps, captured by the force balance but not by the base suction that only takes into account the pressure on the vertical base. The side force variation in Fig.~\ref{fig:Comp_TB}($c$) remains quite similar, in agreement with the absence of significant effect on the horizontal base pressure gradient (Fig.~\ref{fig:Comp_HF_GYGZ}($a$)). As expected, the major effect is on the lift coefficient in Fig.~\ref{fig:Comp_TB}($d$) in the intermediate yaw range where the vertical base pressure gradient is the most affected by the torsion (Fig.~\ref{fig:Comp_HF_GYGZ}($b$)). Different vertical base pressure gradient indicate different flaps deformations associated with modified lift coefficients.

\begin{figure}[t]
\centering
\includegraphics[width=0.75\textwidth]{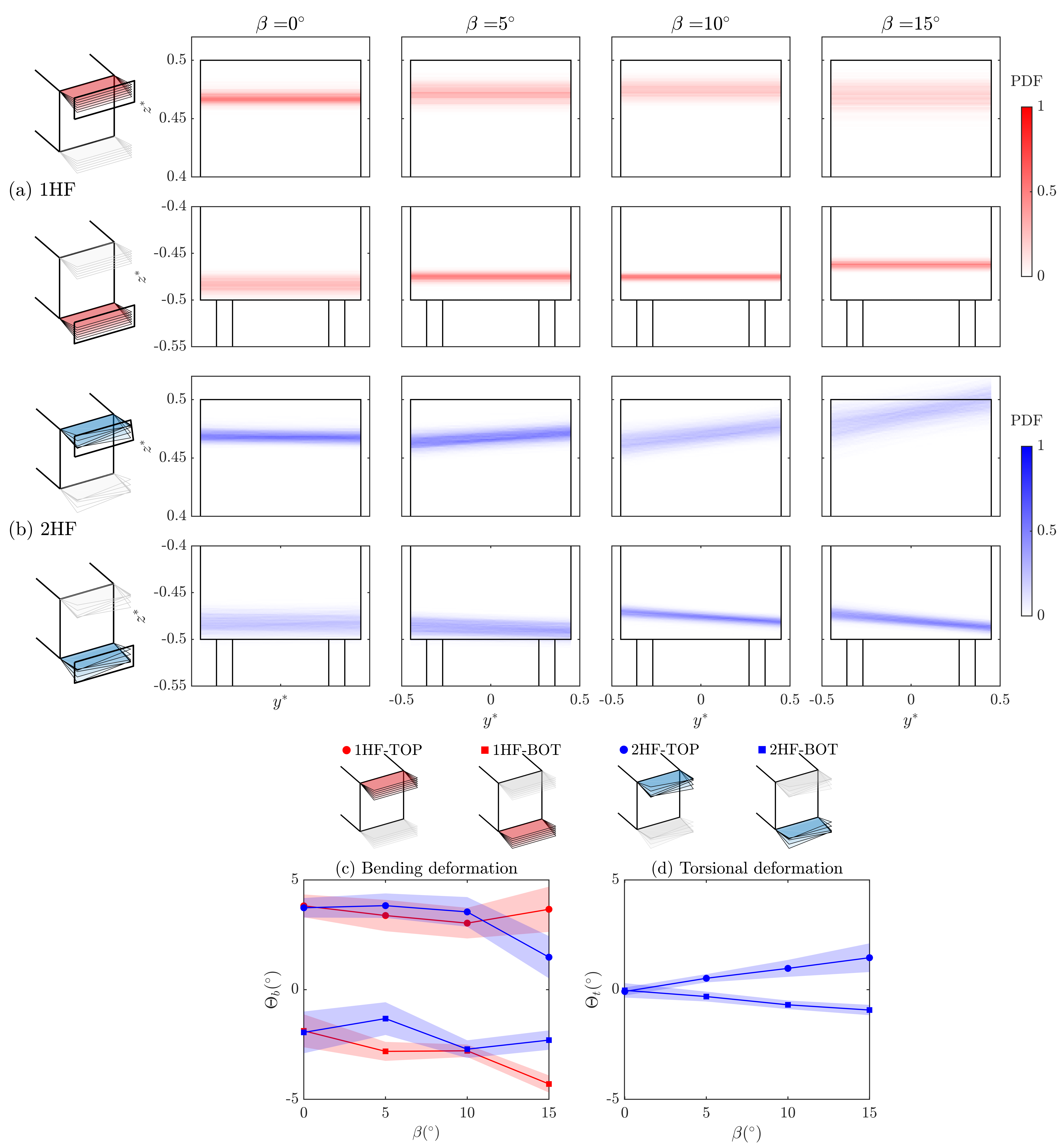}
\caption{\label{fig:FlapOrientation} Contours of the probability density function (PDF) of the flap trailing edge position for $\beta=[0:5:10:15]^{\circ}$ at $Re= 2.13 \cdot 10^5$ for (a) 1HF (top row, red scale) and (b) 2HF (bottom row, blue scale) configurations. Variation of (c) bending and (d) torsion angles versus yaw $\beta$, at $Re= 2.13 \cdot 10^5$. The average values, $\Theta$, for the top flap are shown with circles and those for the bottom flap with squares. The amplitude of the flap deformation, $\Theta ^{\prime}$,  is depicted with shadows.}
\end{figure}

The flaps deformations are shown in Fig.~\ref{fig:FlapOrientation} for different yaws, where  Fig.~\ref{fig:FlapOrientation}($a$) illustrates the case with bending only flaps (1HF) and  Fig.~\ref{fig:FlapOrientation}($b$), the case with bending/torsion flaps (2HF). The PDF contour of the flap trailing edge position allows to evaluate both the static and fluctuating deformation of the flaps. The next Figs.~\ref{fig:FlapOrientation}($c,d$) provides the same information but quantifies the bending angle $\Theta_b$ and twist angle $\Theta_t$ of the flap deformation as defined in Figs.~\ref{fig:SetUp}($d,e$). We first comment on the static bending deformation of top and bottom flaps having a major impact on the lift coefficient. From the study of \cite{Grandemange_2013-ExiF} using a very similar geometry at zero yaw, the boat tail effect produced by the flap deflections is responsible for the drag decrease (see also previous studies from \cite{Mair1978,Wong_1983}) and the lift variation from horizontal rigid flap is linear with the top and bottom bending angles of the flaps. The positive lift variation observed for both 1HF and 2HF around zero yaw in Fig.~\ref{fig:Comp_TB}($d$) is due to the large positive deflection of the top angle (see Fig.~\ref{fig:FlapOrientation}($c$)). Around $\beta=5^\circ$, the vanishing lift variation of the 1HF configuration seems to be due to the large negative deflection of the bottom angle that amplifies at $\beta=15^\circ$ in agreement with the negative lift variation. For the 2HF case, the reduction of the top bending angle when yaw is increased seems to be in agreement with the lift decrease to zero. The static twist angle for the 2HF configuration of the top and bottom flaps increases with yaw in an antisymmetric manner such that the vertical separating distance between the top and bottom flap is larger on the windward side than on the leeward side as clearly observable in Fig.~\ref{fig:FlapOrientation}($b$) when yaw increases. In the view of the cross-flow, the static twist deformation shows some sort of a boat tailing arrangement. The flap deformations can have very large fluctuations as reported by the shadow representation in Fig.~\ref{fig:FlapOrientation}. The unsteadiness significantly changes with yaw; for both the bending and twist angles the fluctuation increases with yaw for the top flap but decreases for the bottom flap. Hence at large yaw, the top flap deformation is much more unsteady than the bottom flap, while the opposite is observed at no yaw.

\begin{figure}[t]
\centering
\includegraphics[width=0.85\textwidth]{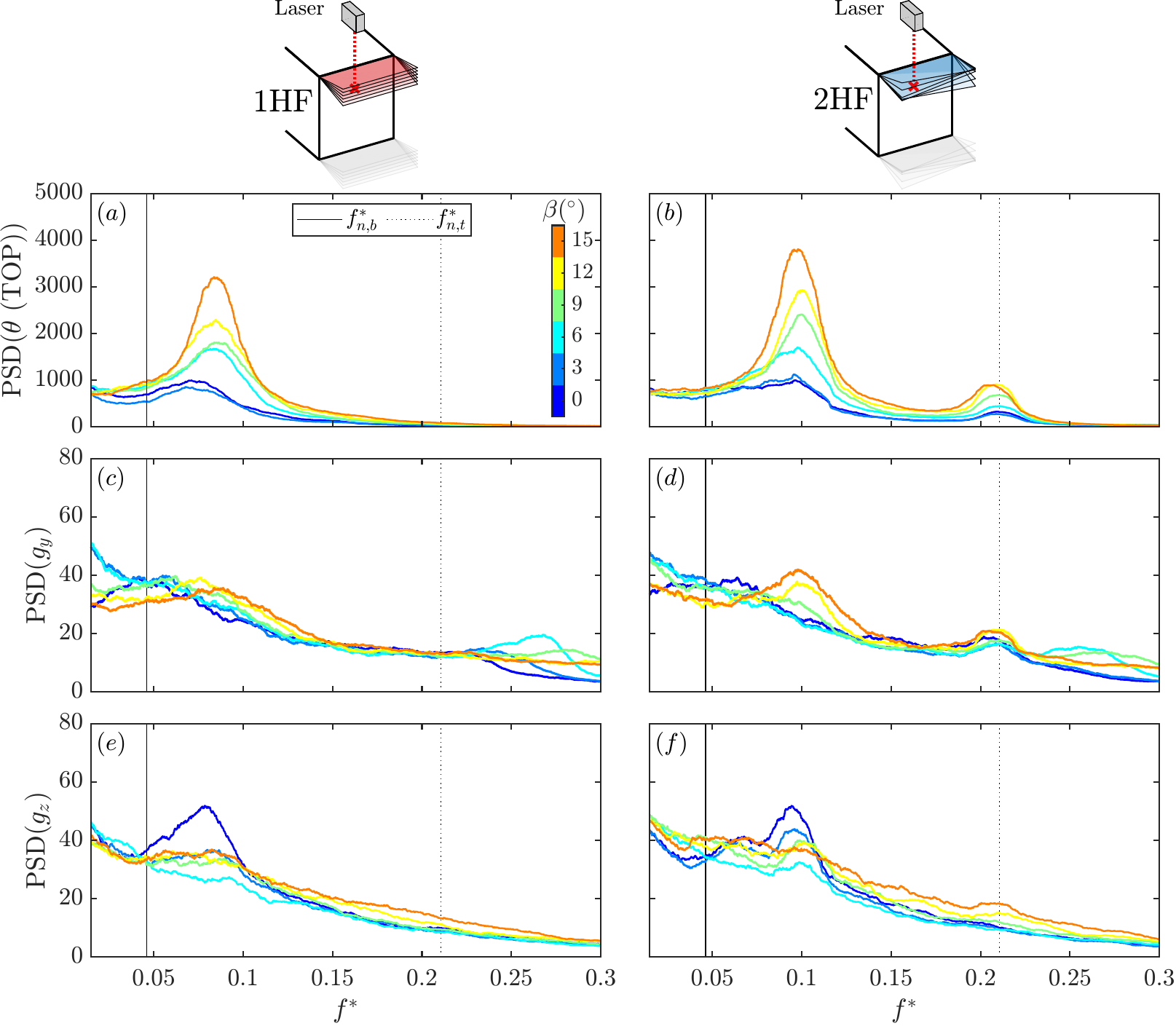}
\caption{\label{fig:PSDCoupling} ($a,b$) Power Spectral Density (PSD) of the vertical displacement of the top flap at $x^{*}= 0.25$, $y^{*} = -0.225$. ($c-f$) PSD of the horizontal, $g_y$, and vertical, $g_z$, base pressure gradient components for 1HF and 2HF configurations for the tested yaw angles at $Re= 2.13 \cdot 10^5$. The bending ($f_{n,b}^*$, black dotted line) and torsion ($f_{n,t}^*$, black dashed line) natural frequencies obtained with no wind are depicted too. }
\end{figure}
The nature of the flap fluctuations is studied using a laser displacement measurement having a better time resolution of 1000~Hz than the $100$~Hz of the stereo PIV system that was used to obtain Fig.~\ref{fig:FlapOrientation}. Due to optical access restriction in the test section, only the top flap motion is recorded with the laser pointing at the left hand-side trailing edge of the flap as shown in the first row of Fig.~\ref{fig:PSDCoupling}. For the 1HF configuration, only the bending motion is captured in Fig.~\ref{fig:PSDCoupling}($a$) that shows a wide peak suggesting a single mode frequency. Whatever the yaw angle, the peak frequency is always larger than the natural frequency obtained with no wind (displayed as the vertical line at $f^*= f*h/u_{\infty}\approx 0.05$). This observation is a classical aeroelastic coupling of added aerodynamic stiffness produced by the flap alignment with the wind, such as a weather vane effect. It can be seen that in the presence of wind, the frequency increases with yaw angle. It is plausibly an effect of a different aerodynamic load with yaw, with a greater amplitude accordingly to the increased frequency peak energy. For the 2HF configuration, bending motion is similarly captured in Fig.~\ref{fig:PSDCoupling}($b$) but at slightly higher frequencies and the twist motion appears in a second broad peak very close to the natural torsion frequency obtained with no wind (displayed as the vertical line at at $f^*\approx 0.21$). There is no observable added aerodynamic stiffness for the aeroelastic coupling about the torsion deformation.

Regarding the base flow, here evaluated with the PSD of the two components of the base pressure gradient in Fig.~\ref{fig:PSDCoupling}($c,e$) for 1HF and Fig.~\ref{fig:PSDCoupling}($d,f$) for 2HF, it seems that $g_z$ is very receptive to the bending motion at small yaw only as indicated by the dark blue peaks in Figs.~\ref{fig:PSDCoupling}($e,f$), while $g_y$ is more responsive at large yaw, but that also corresponds to larger bending excitations. The twist frequency only appears significantly in $g_y$ in Fig.~\ref{fig:PSDCoupling}($f$). The dynamical effect on the base pressure distribution is assumed to be a consequence of the motion of the separation line imposed at the flap trailing edge (the flap deflection angles are too small to consider any anticipated smooth separation). The change of the recirculating bubble shape with the flaps motion likely corresponds to the global flow response capable of modifying the base pressure distribution.

\begin{figure}[t]
\centering
\includegraphics[width=0.75\textwidth]{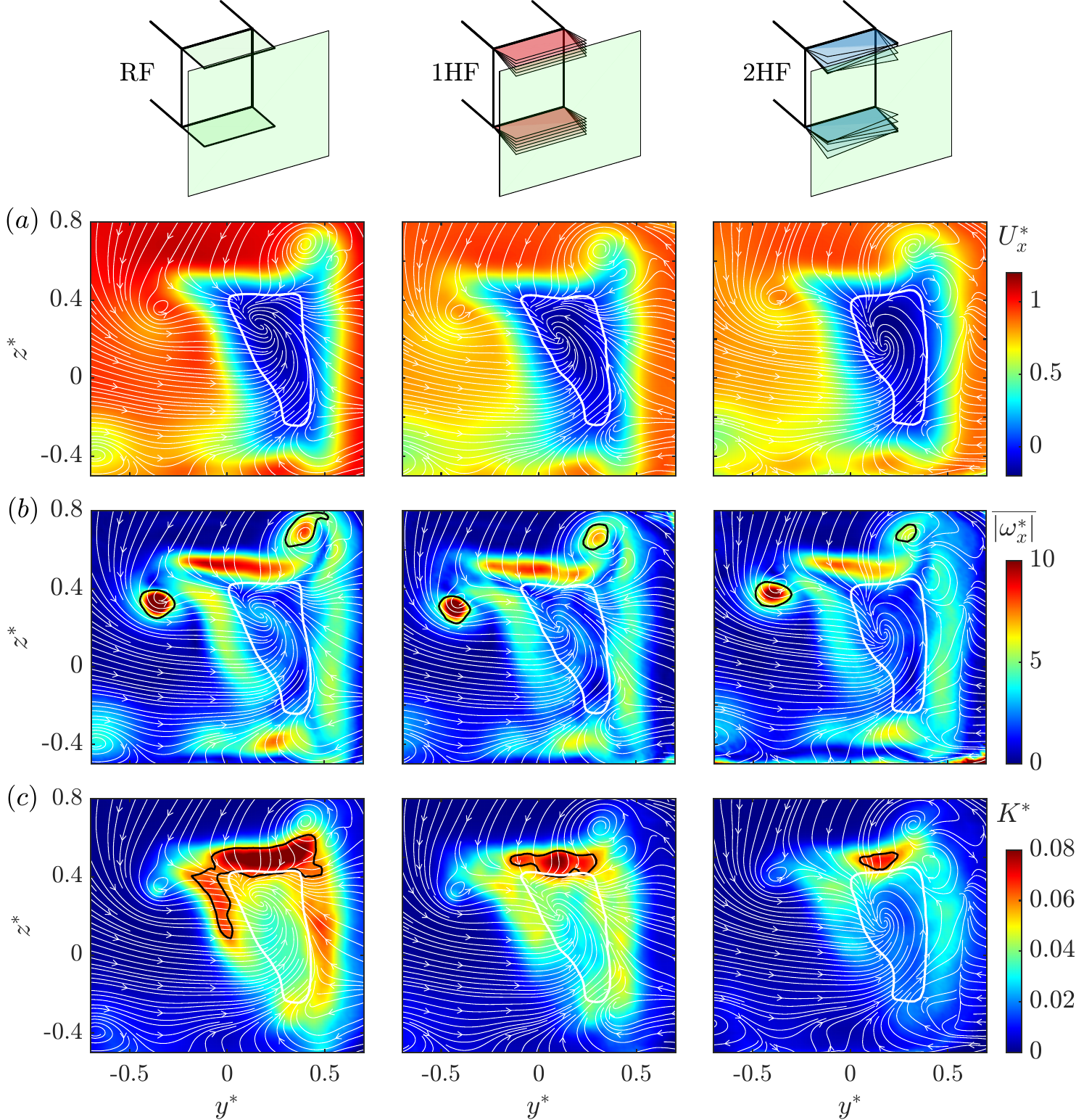}
\caption{\label{fig:PIV_beta15}Time-averaged contours of ($a$) streamwise velocity, $U_x^*$, ($b$) magnitude of streamwise vorticity, $\overline{|\omega_x^*|}$ , and ($c$) turbulent kinetic energy, $K^*$, at $x^*=0.915$ for $\beta = 15^\circ$ and $Re= 2.13 \cdot 10^5$ for the RF, 1HF and 2HF configurations. Thin white lines illustrate the flow streamlines ($U^{*}_{y}$, $U^{*}_{z}$), while thick white line shows the isoline of $U_x^*=0$. In ($b$), the black line represents the isoline $\overline{|\omega_x^*|}=5$  of the longitudinal vortices occurring on the leeward and windward side while ($c$) limits the iso-contour of $K^* = 0.06$. }
\end{figure}

The velocity field of the mean wake is shown in Fig.~\ref{fig:PIV_beta15} for the crosswind condition where the twist deformation of the 2HF flaps is the most significant ($\beta = 15^{\circ}$). The presence of a back-flow ($U^*<0$) in the recirculating bubble of the three cases in Fig.~\ref{fig:PIV_beta15}($a$), compared to the baseline in Fig.~\ref{fig:BasePIV}($a$) at same yaw for which no back-flow was observed, is an indication of an extended bubble length. The larger distance between the bubble closure and the base of the body indicates globally less curved streamlines in the larger speed regions, reducing centripetal acceleration and the associated pressure gradient which results in a higher base pressure. It is likely that a contribution to the drag reduction with the flaps is caused by the same basic effect as for a deep cavity at the rear of the Ahmed body \citep{Evrard16,Lucas17}.

The mean of the streamwise vorticity modulus in Fig.~\ref{fig:PIV_beta15}($b$) shows a gradual decrease of its maximum from RF to 2HF. When concentrated in circular shapes, the corresponding coherent structures are longitudinal vortices, sources of low pressure that contribute to drag; their intensity can be characterised by their circulation. The circulation can be assessed by choosing arbitrarily a contour at $\overline{|\omega_{xV}^*|}=5$ delimiting a surface $S_V^*$ of the vortex providing a circulation $\Gamma_V^*=\iint_{S_V} \overline{|\omega_{xV}^*|}\times dS^* \approx \langle \overline{|\omega_{xV}^*|}\rangle_{S_V}\times S_V^*$, where $\langle \overline{|\omega_{xV}^*|}\rangle_{S_V}$ denotes the average vorticity modulus on the vortex surface. The contours at $\overline{|\omega_{xV}^*|}=5$ are shown as continuous black lines in Fig.~\ref{fig:PIV_beta15}($b$). Table~\ref{tab:Vortex-intensities} recaps the obtained intensities for the longitudinal vortex identified in the wake of the baseline in Fig.~\ref{fig:BasePIV}($b$) and the three cases in Fig.~\ref{fig:PIV_beta15}($b$).
The first observation is an intensification of the vortices when rigid flaps are added. While the windward vortex (WW) was not observable for the baseline in Fig.~\ref{fig:BasePIV}($b$) it is now clearly formed in  Fig.~\ref{fig:PIV_beta15}($b$), and both the vorticity maximum and circulation are slightly increased on the leeward (LW) side (see Table~\ref{tab:Vortex-intensities}). Using deformable flaps decreases their intensities, with the better attenuation for the windward vortex where the 1HF and 2HF flaps achieve reductions in terms of vortex circulation, $\Gamma_V^*$, of 25.2\% and 30\% in the leeward vortex, and of 43\% and 62.8\% in the windward vortex, respectively, compared with the RF case, as reported in Table~\ref{tab:Vortex-intensities}. Since these vortices are associated with induced drag, their attenuation offer a plausible explanation for the additional drag reduction as induced drag reduction compared to the RF configuration. However, this is probably not the only cause, as the elastic flaps present significant vibrations at yaw that might control the development of the separating mixing layers (see \cite{Greenblatt_2000} and references therein). The turbulent kinetic energy in the wake shown in Fig.~\ref{fig:BasePIV}($c$) indicates a spectacular stabilisation of the mixing layer using the elastic flaps, especially the 2HF ones. This strong effect that attenuates the mixing layer spreading rate might also modify the recirculating bubble shape to contribute to the base pressure increase.

% \begin{table}[]
% \centering
% \begin{tabular}{c|cc|cc|cc|cc}
% Configuration&~~~~B&&~~~~RF&&~~~~1HF&&~~~~2HF&\\
% Vortex&LW&WW&LW&WW&LW&WW&LW&WW\\ \hline
% $\overline{|\omega_{xV}^*|}_\text{Max}$&15.97&...&18.38&\tcr{?}&14.71&\tcr{?}&11.45&\tcr{?}\\
% $\Gamma_V^*$&0.095&...&0.103&0.084&0.077&0.050&0.072&0.032\\
% \end{tabular}
% \caption{Intensities of the longitudinal vortices identified on the leeward (LW) and windward (WW) hand side of the wake at $\beta=15^\circ$ in Fig.~\ref{fig:BasePIV}($b$) and Fig.~\ref{fig:PIV_beta15}($b$) obtained for the baseline (B) and the 3 TB configurations (RF, 1HF, 2HF). Intensities are characterised as the local maximum of the mean vorticity modulus, $\overline{|\omega_{xV}^*|}_\text{Max}$ and vortex circulation, $\Gamma_V^*$ (see text).}
% \label{tab:Vortex-intensities}
% \end{table}

\begin{table}[]
\centering
\begin{tabular}{c|cc|cc|cc|cc}
Configuration                           & \multicolumn{2}{c|}{B} & \multicolumn{2}{c|}{RF} & \multicolumn{2}{c|}{1HF} & \multicolumn{2}{c}{2HF} \\
Vortex                                  & LW          & WW       & LW         & WW         & LW          & WW         & LW         & WW         \\ \hline
$\overline{|\omega_{xV}^*|}_\text{Max}$ & 15.97       & --       & 18.38      & 8.43       & 14.71       & 7.311      & 11.45      & 6.00       \\
$\Gamma_V^*$                            & 0.095       & --       & 0.103      & 0.086      & 0.077       & 0.049      & 0.072      & 0.032    
\end{tabular}
\caption{Intensities of the longitudinal vortices identified on the leeward (LW) and windward (WW) hand-side of the wake at $\beta=15^\circ$ in Fig.~\ref{fig:BasePIV}($b$) and Fig.~\ref{fig:PIV_beta15}($b$) obtained for the baseline (B) and the 3 TB configurations (RF, 1HF, 2HF). Intensities are characterised as the local maximum of the mean vorticity modulus, $\overline{|\omega_{xV}^*|}_\text{Max}$ and vortex circulation, $\Gamma_V^*$ (see text).}
\label{tab:Vortex-intensities}
\end{table}

%% Influence of reduced velocity with contourst, references to PRF and JFS here
\subsubsection{Influence of the flexibility of the flaps}\label{subsubsec:Ur}
\begin{figure}[t]
\centering
\includegraphics[width=0.8\textwidth]{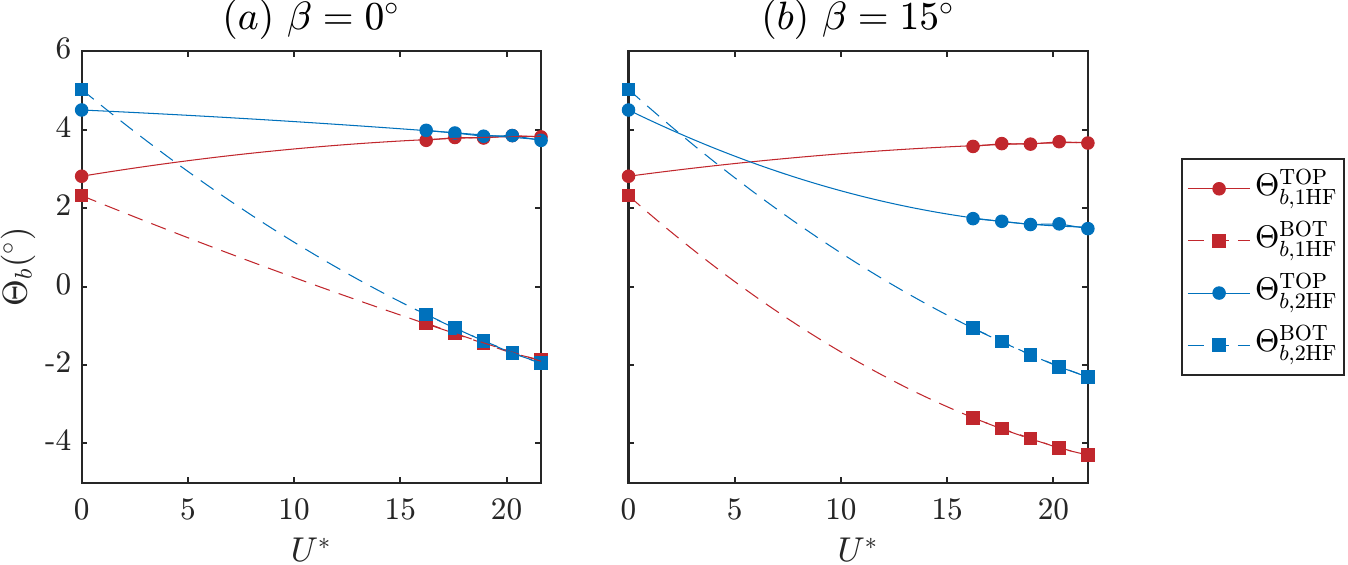}
\caption{\label{fig:AeroReEffect} Static flap deformation in bending, $\Theta_b$ vs. reduced velocity $U^{*}$ at $\beta = 0^{\circ}$ ($a$) and $15^{\circ}$ ($b$) for the Top-Bottom (TB) arrangement in the 1HF (red symbols) and 2HF (blue symbols) configurations.}
\end{figure}

The self-adaptive elastic flaps deformation depends on the velocity ratio between the characteristic flap motion and the incoming flow, this ratio is usually expressed in terms of the reduced velocity, chosen here as  $U^{*} = u_{\infty}/f_{n,b}h$. As the reduced velocity increases, the flaps are relatively more flexible and better adapt to the incoming flow \citep{CamachoSanchez23}, however, too flexible flaps may undergo fluttering \citep{MunozHervas2024}, having a negative impact on the aerodynamics of body. In this work, the effect of varying $U^*$ has also been studied, Figure \ref{fig:AeroReEffect}($a$) shows the top and bottom static bending deformation of the 1HF (red) and 2HF (blue) flaps at $\beta=0^\circ$ against the reduced velocity $U^*$ (note that here $U^*$ is modified by changing the freestream velocity $u_{\infty}$). With no wind, the static deformation due to gravity is responsible for the initial deflection of approximately $3^\circ$ for the 1HF flaps and  $5^\circ$ for the 2HF flaps. Despite the different initial conditions, both the 1HF and the 2HF tend to same deflections at the largest reduced velocity ($U^*=21.6$) corresponding to results shown in Fig.~\ref{fig:FlapOrientation}, and employed in the rest of the paper, for the $\beta=0^\circ$ case. This is totally different at yaw as shown for $\beta=15^\circ$ in Fig.~\ref{fig:AeroReEffect}($b$) where the deflections obtained for the largest reduced velocity corresponds to results shown in Fig.~\ref{fig:FlapOrientation} at the same yaw. The different bending deformation obtained for the 2HF compared to the 1HF flaps is assumed to be related to the additional torsional degree of freedom of the 2HF flaps and not exclusively to the different initial static deflections at $U^{*} = 0$, otherwise different deformations would have been expected in Fig.~\ref{fig:AeroReEffect}($a$) at the largest reduced velocity. %\tcr{anything else to say ?}
\section{Conclusions}
\label{sec:Conclusions}
This study investigates the potential for drag-reduction of low-mechanical-order, self-adaptive control devices in three-dimensional wakes, as an alternative to purely flexible appendages, that may undergo unstable dynamic responses.
In particular, the ability of rear additional rigid panels to reduce drag under yaw conditions has been investigated in a wind tunnel on a taller-than-wide square-back Ahmed body. %Different arrangements are studied in order to compare the potential for drag-reduction of low-mechanical-order, self-adaptive control systems. 
Thus, the pair of panels is, depending on the their arrangement, either placed along the vertical (LR) or the horizontal (TB) edges of the base. When the panels are fixed rigidly to the sides of the model, both arrangements (LR and TB) reduce the drag at no yaw. The origin of drag reduction is found to be similar to a rear cavity effect, that elongates the recirculation area taken between the body base and the recirculating flow closure thus reducing the base suction. The steady instability, related to the RSB mode, that appears in the vertical direction for this geometry, is also found to be suppressed. While both arrangements are equivalent to reduce the drag at no yaw, their behaviour differs drastically at yaw. The vertical arrangement (LR) presents an obvious increasing obstruction to the cross flow as the yaw is increased such that the pair of panels eventually increases the drag for yaw larger than 3$^\circ$. 
There is a major improvement by making their fixation flexible (LR-1HF), and the angles adaptation leads satisfactorily to drag reduction at any yaw, with even better performance at no yaw due to the boat-tailing shape of the panels. However, drag reduction always remains smaller than that obtained with the straight rigidly fixed pair of horizontal panels (TB) at any yaw larger than 3$^\circ$. A major conclusion is that extended vertical panels (LR) placed at the rear of a rectangular base might not be the best drag reducer appendage, even with adaptive shape, in real road conditions where atmospheric crosswinds are unavoidable. For applications, it is speculated that horizontal appendages (TB) might have better performances in terms of a wind-averaged drag coefficient (see \citealt{Cooper_1976, Ingram_1978, Howell_2015}), taking into account crosswinds. The wind-averaged drag coefficient, as introduced in \citep{Howell_2018,Varney_2018} for typical European wind conditions, is calculated for all arrangements of the present study in Table~\ref{tab:WindAvgDrag}. We can see in table that, indeed, horizontal appendages (TB) are always better to reduce the wind averaged drag coefficient whether the fixation  is rigid (RF) or elastic (1HF and 2HF).

The horizontal pair of panels (TB) sustains the cavity effect at any yaw without producing the extra drag related to the cross flow as discussed above. However, the three-dimensional wake structure reveals a clear intensification of the longitudinal vortices on both the leeward and windward sides emanating from the body three-dimensional separations at yaw. This detrimental effect of the horizontal panels, source of additional induced drag, is mitigated with the elastic fixations. The static deformations of the panels correspond to a boat tailing, both in the bending and the torsion direction.

It is also found that as the yaw is increased, the elastic fixation reduces the turbulent fluctuation in the wake while the vibration amplitude of the panels increases. This apparent contradiction raises the possibility of an active-control role for this vibration on the mixing layer developments. To conclude, the novel elastic fixation allowing bending and torsion of the panels achieves in variable yaw conditions, ($i$) the best wind-averaged drag reduction, ($ii$) longitudinal vortices mitigation and ($iii$) turbulent kinetic energy reduction. \\

\begin{table}[]
\centering
\begin{tabular}{c|c|c|c}
\hline
\#                   & Arrangement & $C_{DWC}$ & $\Delta C_{DWC}$ (\%) \\ \hline
B                    & -         & 0.395     & -                     \\ \hline
\multirow{2}{*}{RF}  & TB        & 0.372     & -5.76                 \\
                     & LR        & 0.405     & 2.53                  \\ \hline
\multirow{2}{*}{1HF} & TB        & 0.370     & -6.32                 \\
                     & LR        & 0.371     & -5.95                 \\ \hline
2HF                  & TB        & 0.365     & -7.62                 \\ \hline
\end{tabular}
\caption{Wind-averaged drag coefficient, $C_{\text{DWC}} = 0.53 C_{x_{\beta=0^\circ}} + 0.34 C_{x_{\beta=5^\circ}}+ 0.13 C_{x_{\beta=10^\circ}}$ as defined in  \cite{Varney_2018}, and the corresponding variation with respect to the baseline case, $\Delta C_{DWC}$.}
\label{tab:WindAvgDrag}
\end{table}

\textbf{Acknowledgments}\\
The authors gratefully acknowledge the funding provided by the projects TED2021-131805B-C21, TED2021-131805B-C22 and PID2022-140433NA-I00 financed by the Spanish MCIN/ AEI/10.13039/501100011033/,  the European Union NextGenerationEU/PRTR and FEDER, UE respectively. J.M.C.S.
want to thank the University Jaén Doctoral School for the financial support provided for the research stay in the University of Liverpool.  M.L.D. also acknowledges the grant  RYC2023-044496-I financed by MICIU/AEI /10.13039/501100011033 and FSE+. This work has been supported by the Khalifa University of Science and Technology under Award No. RIG-2023-024.
%\end{acknowledgments}

%\appendix

%\section{Appendix}
%\label{sec:appendix}

 % \bibliographystyle{elsarticle-num} 
 % \bibliography{cas-refs}
 
\bibliographystyle{elsarticle-harv}
\bibliography{cas-refs}





% \end{thebibliography}
\end{document}